%% file: main.tex
\documentclass[format=acmsmall, review=false, screen=false]{acmart}
\usepackage{comment}
\usepackage{todonotes}

\usepackage{subcaption} 
\usepackage{caption}  
\usepackage{sidecap}
\usepackage[utf8]{inputenc}
\usepackage[inline]{enumitem}

\DeclareCaptionFormat{myformat}{\fontsize{9}{9}\selectfont#1#2#3}
\sidecaptionvpos{figure}{c}
\usepackage{textcomp} 

\usepackage[ampersand]{easylist}
\usepackage{amsmath}
\usepackage{amsthm}
\usepackage{makecell}
\newtheoremstyle{mystyle}
{}
{}
{}
{}
{\itshape}
{.}
{ }
{}
\newtheoremstyle{mystyle2}
{}
{}
{\itshape}
{}
{\scshape}
{.}
{ }
{}
\theoremstyle{acmplain}

\theoremstyle{acmplain}
\usepackage[T1]{fontenc}

\usepackage{graphicx}

\usepackage{listings}

\usepackage{caption}
\usepackage{subcaption}
\usepackage{mathtools}

\usepackage[]{units}
 
\usepackage{multirow} 
\usepackage{lipsum}
\usepackage{xcolor,colortbl}
\usepackage{tikz}
\usetikzlibrary{positioning,  math, shadings}
\usetikzlibrary{calc,decorations.pathreplacing}
\usetikzlibrary{patterns}
\usetikzlibrary{arrows,automata, shapes, petri}
\usetikzlibrary{arrows.meta, shapes.callouts}

\input{fig/defs.tex}

\usetikzlibrary{shapes.symbols} 
\usetikzlibrary{shapes.geometric} 
\tikzset{
  hexagon/.style={signal,signal to=east and west}
}
\tikzset{
  octagon/.style={shape=regular polygon, regular polygon sides=8, draw, minimum width=.8in}
}

\usepackage{pgfplots}
\usepackage{booktabs}

\usepackage{soul}
\usepackage{hyperref}
\hypersetup{%
	colorlinks=false,
	linkbordercolor=red,
	pdfborderstyle={/S/U/W 1}
}

\newcolumntype{L}[1]{>{\raggedright\let\newline\\\arraybackslash\hspace{0pt}}m{#1}}
\newcolumntype{C}[1]{>{\centering\let\newline\\\arraybackslash\hspace{0pt}}m{#1}}
\newcolumntype{R}[1]{>{\raggedleft\let\newline\\\arraybackslash\hspace{0pt}}m{#1}}

\usetikzlibrary{decorations.pathreplacing, patterns}

\usepackage{algorithm} 
\usepackage{algorithmic}
\usepackage{diagbox} 
\input{defs.tex}

\usepackage{filecontents}

\begin{document}

\title{Ratatoskr: An open-source framework for in-depth power, performance and area analysis in 3D NoCs}

\author{Jan Moritz Joseph}
\affiliation{%
	\institution{Otto-von-Guericke-Universit\"at Magdeburg}
	\streetaddress{Univerist\"atsplatz 2}
	\city{Magdeburg}
	\state{Sachsen-Anhalt}
	\postcode{39106}
	\country{Germany}}
\email{moritz.joseph@ovgu.de}

\author{Lennart Bamberg}
\affiliation{%
	\institution{Universit\"at Bremen}
	\streetaddress{Otto-Hahn Allee 1}
	\city{Bremen}
	\state{Bremen}
	\postcode{28359}
	\country{Germany}}
\email{bamberg@item.uni-bremen.de}

\author{Imad Hajjar}
\affiliation{%
	\institution{Otto-von-Guericke-Universit\"at Magdeburg}
	\streetaddress{Univerist\"atsplatz 2}
	\city{Magdeburg}
	\state{Sachsen-Anhalt}
	\postcode{39106}
	\country{Germany}}
\email{imad.hajjar@st.ovgu.de}

\author{Behnam Razi Perjikolaei}
\affiliation{%
	\institution{Universit\"at Bremen}
	\streetaddress{Otto-Hahn Allee 1}
	\city{Bremen}
	\state{Bremen}
	\postcode{28359}
	\country{Germany}}
\email{raziperj@uni-bremen.de}

\author{Anna Drewes}
\affiliation{%
	\institution{Otto-von-Guericke-Universit\"at Magdeburg}
	\streetaddress{Univerist\"atsplatz 2}
	\city{Magdeburg}
	\state{Sachsen-Anhalt}
	\postcode{39106}
	\country{Germany}}
\email{anna.drewes@st.ovgu.de}

\author{Alberto Garc\'ia-Ortiz}
\affiliation{%
	\institution{Universit\"at Bremen}
	\streetaddress{Otto-Hahn Allee 1}
	\city{Bremen}
	\state{Bremen}
	\postcode{28359}
	\country{Germany}}
\email{agarcia@item.uni-bremen.de}

\author{Thilo Pionteck}
\affiliation{%
	\institution{Otto-von-Guericke-Universit\"at Magdeburg}
	\streetaddress{Univerist\"atsplatz 2}
	\city{Magdeburg}
	\state{Sachsen-Anhalt}
	\postcode{39106}
	\country{Germany}}
\email{thilo.pionteck@ovgu.de}

\renewcommand{\shortauthors}{J.M. Joseph et al.}

\begin{abstract}
	\input{chapters/00_abstract.tex}
\end{abstract}


\maketitle

\input{chapters/01_introduction.tex}

\input{chapters/02_related.tex}
\input{chapters/03_methods.tex}
\input{chapters/04_results.tex}
\input{chapters/05_conclusion.tex}

\begin{acks}
This work is funded by the German Research Foundation (DFG) projects PI 447/8 and GA 763/7. 
\end{acks}

\bibliographystyle{ACM-Reference-Format}
\bibliography{bib}

\end{document}

%% file: fig/defs.tex
\tikzset{router/.style={circle,draw,fill=gray!60,inner sep=0pt,minimum size=5pt}}
\usetikzlibrary{fadings}
\tikzset{desc/.style={font = \footnotesize\sffamily}}

\definecolor{col1}{RGB}{27,158,119}
\definecolor{col2}{RGB}{217,95,2}
\definecolor{col3}{RGB}{117,112,179}


%% file: defs.tex
\DeclareMathAlphabet{\mymathbb}{U}{BOONDOX-ds}{m}{n} 

% ...

\lstdefinestyle{routingScript}{
	emph={%
	if, else, then, route to
	},
    emphstyle={\bfseries},
	morekeywords={if, else, then, route, to, end},
	escapeinside={(*}{*)},
	numbers=left,
	stepnumber=2,
	numbersep=10pt,
	tabsize=2,
	showspaces=false,
	showstringspaces=false
}

\lstdefinestyle{linkmodel}{
	belowcaptionskip=5pt,
	abovecaptionskip=0pt,
	belowskip = 4pt,
	aboveskip = 4pt,
	breaklines=true,
	frame=bt,
	language=Python,
	keywords=[1]{Interconnect, Driver, DataStream, DataStreamProb},
	keywords=[2]{max_metal_wire_length, E, area_3D, from_stoch, predefined},
	keywords=[3]{from, import},
	showstringspaces=false,
	basicstyle=\tiny \ttfamily,
	keywordstyle=[1]\color{col1},
	keywordstyle=[2]\color{col2},
	keywordstyle=[3]\color{col3},
	commentstyle=\itshape\color{col3},
	stringstyle=\color{purple},
	identifierstyle=\color{gray},
	tabsize=2,
	numbersep=8pt,
	stepnumber=1,
	numberstyle=\tiny\color{gray},
}

\lstdefinestyle{myxml}{
	belowcaptionskip=5pt,
	abovecaptionskip=0pt,
	belowskip = 4pt,
	aboveskip = 4pt,
	breaklines=true,
	frame=bt,
	keywords=[1]{nodeTypes, nodeType, layerTypes, layerType, nodes, node, connections, con, ports, port, map, bind, data, dataTypes, dataType, task, requires, requirement, generates, possibility, destinations, destination, synthetic, phase, distribution },
	keywords=[2]{routerModel, routing, selection, clockSpeed, technology, xPos, yPos, zPos, idType, length, width, depth, interface, bufferDepth, vcCount, name, start, duration, repeat, type, souce, count, probability, delay, interval, count, source, hotspot, injectionRate},
	keywords=[3]{id, value, min, max},
	showstringspaces=false,
	basicstyle=\tiny \ttfamily,
	keywordstyle=[1]\color{col1},
	keywordstyle=[2]\color{col2},
	keywordstyle=[3]\color{col3},
	commentstyle=\itshape\color{green},
	stringstyle=\color{purple},
	identifierstyle=\color{gray},
	tabsize=2,
	numbersep=8pt,
	stepnumber=1,
	numberstyle=\tiny\color{gray},
}

\tikzstyle{myMarkerPlot}=[mark size=2.5pt, thick, densely dotted, mark options={solid,draw=black}]

%% file: chapters/00_abstract.tex
In this paper introduce \emph{Ratatoskr}, an open-source framework for in-depth power, performance and area (PPA) analysis in Networks-on-Chips (NoCs) for 3D-integrated heterogeneous System-on-Chips (SoCs). It covers several layers of abstraction by providing a RTL NoC hardware implementation, a cycle-accurate NoC simulator and an application model on transaction level. By this comprehensive approach, \emph{Ratatoskr} can provide the following specific PPA analyses: dynamic power of links can be estimated within 2.4\% accuracy of bit-level simulations while maintaining cycle-accurate simulation speed. Router power is determined from RTL synthesis combined with cycle-accurate simulations. The performance of the whole NoC can be measured both via cycle-accurate and RTL simulations. The functionality of routers can be verified using the hardware model. The overall NoC area is estimated at the RTL using a pre-characterization of units at the gate level. Despite these manifold features, \emph{Ratatoskr} offers easy two-step user interaction: First, a single point-of-entry that allows to set design parameters and second, PPA reports are generated automatically. For both the input and the output, different levels of abstraction can be chosen for high-level rapid network analysis or low-level improvement of architectural details. The proposed NoC models reduce total router power up to 32\% and router area by 3\% in comparison to a conventional standard router. As a forward-thinking and unique feature not found in other NoC PPA-measurement tools, \emph{Ratatoskr} supports heterogeneous 3D integration that is one of the most promising integration paradigms for upcoming SoCs. Thereby, \emph{Ratatoskr} lies the groundwork to design their communication architectures.

%% file: chapters/01_introduction.tex
\section{Introduction}

Networks-on-Chips (NoCs), in which on-chip routers connect components (e.g. processing elements) and transmit data using packets, are one of the most promising communication architectures for state-of-the-art chip designs. NoCs have advantages over conventional approaches such as bus systems: Since the number of components per chip increases constantly, buses become a bottleneck as more and more participants fight for arbitration. NoCs tackle this issue and offer better latency and throughput because of their parallel decentralized nature. This increases system efficiency and thus power consumption. Therefore, integration of NoCs into whole systems has become a vital research topic. Recently, NoCs are found in both the academic designs, e.g. in the Eyeriss CNN-accelerator \cite{Chen.2018} or in the bio-inspired visual attention engine \cite{Kim.2009}, and the industrial designs, e.g. in the 260-core ShenWei processor Sunway SW26010 processor \cite{Dongarra.2016} or AMD's infinity fabric \cite{Lepak.2017}. 

In addition to the prevalent trend towards higher core counts, the performance of systems can be increased using novel integration methods. One very promising option is 3D heterogeneous integration: Vertical 3D interconnects allow to reduce delay and power consumption \cite{Dong.2009}; also tackles fundamental limits of computation by asymptotically reducing computation time from $t$ to $t^{.75}$ \cite{Markov.2014}. Furthermore, the footprint shrinks through stacking because components are divided among layers. Beside these incremental advances, 3D integration allows for one game-changing paradigm: Integration of heterogeneous technologies. This is a fundamental advantage that is efficiently only available using 3D integration.\footnote{2.5D integration allows for heterogeneous integration, as well, but is limited by the rather poor performance of the interconnects though the interposer layer in comparison to a true 3D approach.} In heterogeneous 3D integrated circuits (ICs), dies with varying electrical characteristics and technologies (analog, mixed-signal, logic and memory) are stacked and closely interconnected. The key benefit lies in the possibility to optimize the silicon technology node for the components of each die. Therefore, heterogeneous 3D integration boosts performance, energy efficiency and robustness, and it allows building truly innovative novel architectures for various applications such as high-performance computing, e.g. performance enhancements by interleaving stacking of memory and processing dies \cite{Yu.2013}. This paradigm is as also applied in Intel's new Lakefield architecture using Foveros 3D technology \cite{Intel.2019}. To build fast and efficient communication architectures for these systems, is essential to design NoCs specifically targeting heterogeneous 3D-ICs.

As for all components integrated into a chip, an in-depth analysis of power, performance and area (PPA) of the NoC is imperative to judge its properties. The PPA figures must take heterogeneous 3D integration into account as it effects power (e.g. \cite{Bamberg.2017}) and area and performance due to them varying with technology. Here, we propose \textbf{\emph{ratatoskr}: An open-source framework for PPA analysis of 3D NoCs.} It is the first comprehensive framework to precisely determine PPA for NoCs from gate level to transaction level, i.e. including a router hardware implementation, a cycle-accurate router model and a transaction-level application model. Furthermore, it supports heterogeneous 3D integration. Such a comprehensive framework cannot be found in the literature so far. While individual tools exist, neither are these integrated into one single tool flow nor are these able to cope with heterogeneous 3D integration: Simulators such as Noxim \cite{Fazzino.2008} and Booksim 2.0 \cite{Jiang.} allow for performance evaluation of general-purpose NoCs, but do not include router models for heterogeneous 3D integration. Plus, both simulators do not offer an application model on transaction level to inject real-world based traffic pattern out-of-the-box. There exist models for power consumption of routers, such as ORION \cite{Kahng.2009}, but these do not model heterogeneity and do not provide a model for dynamic power consumption of links \cite[Sec. 2]{Joseph.2018c}. Hardware models of routers enable precise area results, but the existing implementations such as OpenSoCFabric \cite{OpenSoCFabric} or OpenSmart \cite{OpenSmart} are not shipped with a simulator for performance comparison and do not reflect the special properties of heterogeneous 3D integration. Thus, \emph{Ratatoskr} addresses these shortcomings by providing the following specific features:
\begin{itemize}
	\item \textbf{Power estimation:} 
		\begin{itemize}
			\item Dynamic power estimation of routers and links on a cycle-accurate level. 
			\item Accuracy of dynamic link energy estimation is within 1\% of bit-level accurate simulations.
		\end{itemize}
	\item \textbf{Performance models:} 
		\begin{itemize}
			\item Network performance for millions of clock cycles on a cycle-accurate (CA) level using the NoC simulator. 
			\item Timing of routers from synthesis on the gate level. This also enables simulation of the NoC on register-transfer level (RTL).
		\end{itemize}
	\item \textbf{Area analysis:} NoC area from synthesis on gate-level for any standard cell technology.
	\item \textbf{Benchmarks:} 
		\begin{itemize}
			\item Support for realistic application model on transaction-level. 
			\item Conventional synthetic traffic patterns.
		\end{itemize}
	\item \textbf{Heterogeneous 3D Integration:}
		\begin{itemize}
			\item Heterogeneity yields non-purely synchronous systems, since the same circuit in mixed-signal and digital achieve different maximum clock speeds. Therefore, the NoC simulation and router hardware model implement a pseudo-mesochronous router (cf. Ref. \cite{Joseph.2018b}).
			\item Heterogeneity yields different number of routers per layers, since (identical) circuits in mixed-signal and digital have different area. Therefore, the NoC simulation allows for any non-regular network topology via XML configuration files.
			\item With the same approach, NoCs in active interposers can be modeled, as well.
		\end{itemize}
	\item \textbf{Usability:} Single point-of-entry to set design parameters. The design parameters allow for rapid testing of different designs.
	\item \textbf{Reporting:} Automatic generation of detailed reports from network-scale performance to buffer utilization.
	\item \textbf{Open-source:} The source code of the framework is available from \url{https://github.com/jmjos/ratatoskr}.
\end{itemize}

The remainder of this paper is structured as follows: In Section~\ref{sec:related} we give a detailed discussion of related approaches. In Section~\ref{sec:tool}, we introduce the \emph{Ratatoskr} framework by starting from a user perspective and highlight the design parameters that can be set. Next, we dig into the technical details: We explain our models in Section~\ref{sec:models} and our implementation of the core components (RTL NoC model, CA NoC model, link model) in Section~\ref{sec:implementation}. After these rather technical sections, we shift the focus again to the user perspective and show the generated reports, which the framework provides for PPA analysis in Section~\ref{sec:reports}. In Section~\ref{sec:results}, we show the results of our framework focusing on properties that the NoC provides as well as the simulation performance; in Section~\ref{sec:discussion} a discussion follows. Finally, we conclude the paper in Section~\ref{sec:conclusion}.

%% file: chapters/02_related.tex
\section{Related Work}\label{sec:related}

As already argued, the \emph{Ratatoskr} framework is the only comprehensive framework for modeling, simulation and design of 3D NoCs that provides in-depth PPA results. Nonetheless, individual tools already exist for some of \emph{Ratatoskr}'s features, from which we discuss differentiating features here.

For the \emph{cycle-accurate performance models} of NoCs, many simulators have been published. An extensive overview is given in Ref.~\cite[Table II, p. 3]{Catania.2016}. Three of them are currently state-of-the-art and used in many publications:
\begin{itemize}
	\item \emph{Noxim} \cite{Catania.2016} is a NoC simulator implemented in C++ using the class library SystemC. It provides the capability to model conventional 2D and homogeneous 3D NoCs. Furthermore, optical links are included. The simulator measures performance (i.e.\ network, packet and flit latency).
	\item \emph{Booksim 2.0} \cite{Jiang.2013} is a NoC simulator implemented in C++. Similar to Noxim, it also provides the capability to model conventional 2D and homogeneous 3D NoCs. It measures performance similar to Noxim. 
	\item  \emph{Garnet 2.0} \cite{Agarwal.2009} targets a different level of abstraction. It is integrated into the Gem5 full-system simulator and therefore offers very high precision to assess different architectures under real loads. However, Garnet 2.0 does not allow for a fast evaluation of millions of clock cycles due to the naturally slow simulation performance of Gem5.
\end{itemize}
Both the Noxim and the Booksim 2.0 simulator are limited to be applied in heterogeneous 3D integration because the provided cycle-accurate router models do not specifically target their unique features, specifically non-purely synchronous clocked routers and non-regular typologies. Plus, it is not possible to model routers with varying parameters per layer within the same NoC without extensive source code modifications. Therefore, the existing tools cannot be used for performance analysis in NoCs for heterogeneous 3D systems. \emph{Ratatoskr} addresses these shortcomings.

Regarding \emph{power estimation} in NoCs, ORION 3.0 \cite{Kahng.2009} is the state-of-the-art tool for modeling power consumption of router's components. The results can either be used to characterize single routers or they can be included in NoC simulators for high-level (i.e. cycle-accurate) power estimation. While Booksim does not include power-modeling capability, Noxim counts energy-relevant events during simulation \cite[Sec. 5.1]{Catania.2016}. The figures from ORION can be used as input to compute the dynamic router power during a simulation run. Since this is currently the best available option for high-level router power consumption in NoCs, we use the same approach and include Noxim's power model. ORION 3.0 also has a basic link energy model (cf. Ref.~\cite[Eqs. 10, 11]{Kahng.2009}) but it does not account for the pattern-dependent coupling switching effects. Since none of the current NoC simulators include power models for links,  those within \emph{Ratatoskr} therefore extend state-of-the-art, as published in \cite{Joseph.2018c}. Please note, that the effects of pattern-dependent coupling switching are not modeled in any power model of NoC routers and this is, to the best of the authors' knowledge, still an open research issue. The \emph{Ratatoskr} framework also allows for power and area measurements post place and route by using the provided NoC implementation . Although very slow, this is the most accurate here. Thereby, \emph{Ratatoskr} covers the full stack of power estimations and can generate results at different accuracy levels and speeds.

For \emph{area analysis} a hardware implementation of a NoC is required. Among the popular open-source implementations, Stanford university \cite{OpenNoCStanford} provides the implementation of the router architecture from \cite{Becker.2014}. As another example, the OpenSoCFabric project \cite{OpenSoCFabric} is quite popular. It allows for 2D mesh and 2D flattened butterfly topology with wormhole routing and virtual channels. The implementation is written in Chisel and enables a run-through ASIC flow. While the provided router is very advanced and rich of features, it does not provide support for 3D integration. OpenSMART implements a "single-cycle multi-hop asynchronous repeated traversal router" \cite[p. 1]{Kwon.2017}, which is state-of-the-art in terms of performance. It allows for non-regular network topologies, but it does not feature 3D integration. Due to their shortcomings, none of the popular NoC implementations can be used for our framework. 

In general, when building a \emph{NoC for heterogeneous 3D integration}, routers must be able to provide two features: First, non-purely synchronous communication must be possible. In Ref.~\cite{Joseph.2018b}, we discuss this topic in detail and propose a pseudo-mesochronous router architecture that enhances throughput by 2$\times$, latency by 2.26$\times$ and dynamic power consumption by 41\% with a small hardware overhead of 2.01\% for synthetic and real-world-based benchmarks using 15nm digital and 45nm mixed-signal technology nodes compared to a conventional NoC design with homogeneous routers, i.e. the same routers regardless of the layer's technology node. Second, reductions of area costs in the routers in mixed-signal layers are essential. This can be achieved by moving resources from this layer into faster and more area-efficient layers. We assess this for buffer re-ordering in Ref.~\cite{Joseph.2017} and achieve area reduction of 28\% and power savings of 15\% at a small 4.6\% performance loss for 130nm and 65nm nodes; for routing algorithms, we show up to 6.5$\times$ latency improvements in Ref.~\cite{Joseph.2018b} for different technology scenarios. These two main previous works form the foundation, on which we develop the router model within the \emph{Ratatoskr} framework. The \emph{Ratatoskr} framework was used to generate results for many publications, e.g. the aforementioned and Ref.~\cite{Bamberg.2017, Bamberg.2018, Bamberg.2018b}. 

%% file: chapters/03_methods.tex
\section{The \emph{Ratatoskr} framework}\label{sec:tool}

Here, we introduce the \emph{Ratatoskr} framework from the user perspective before we gradually go into technical details in the subsequent sections. In this section, we explain the framework's parts and their functionality to generate an in-depth PPA analysis and introduce the process to \emph{set design parameters}. 

\subsection{Parts and functionality of the framework}

\begin{figure}
	\centering
	\includegraphics[width=.73\linewidth]{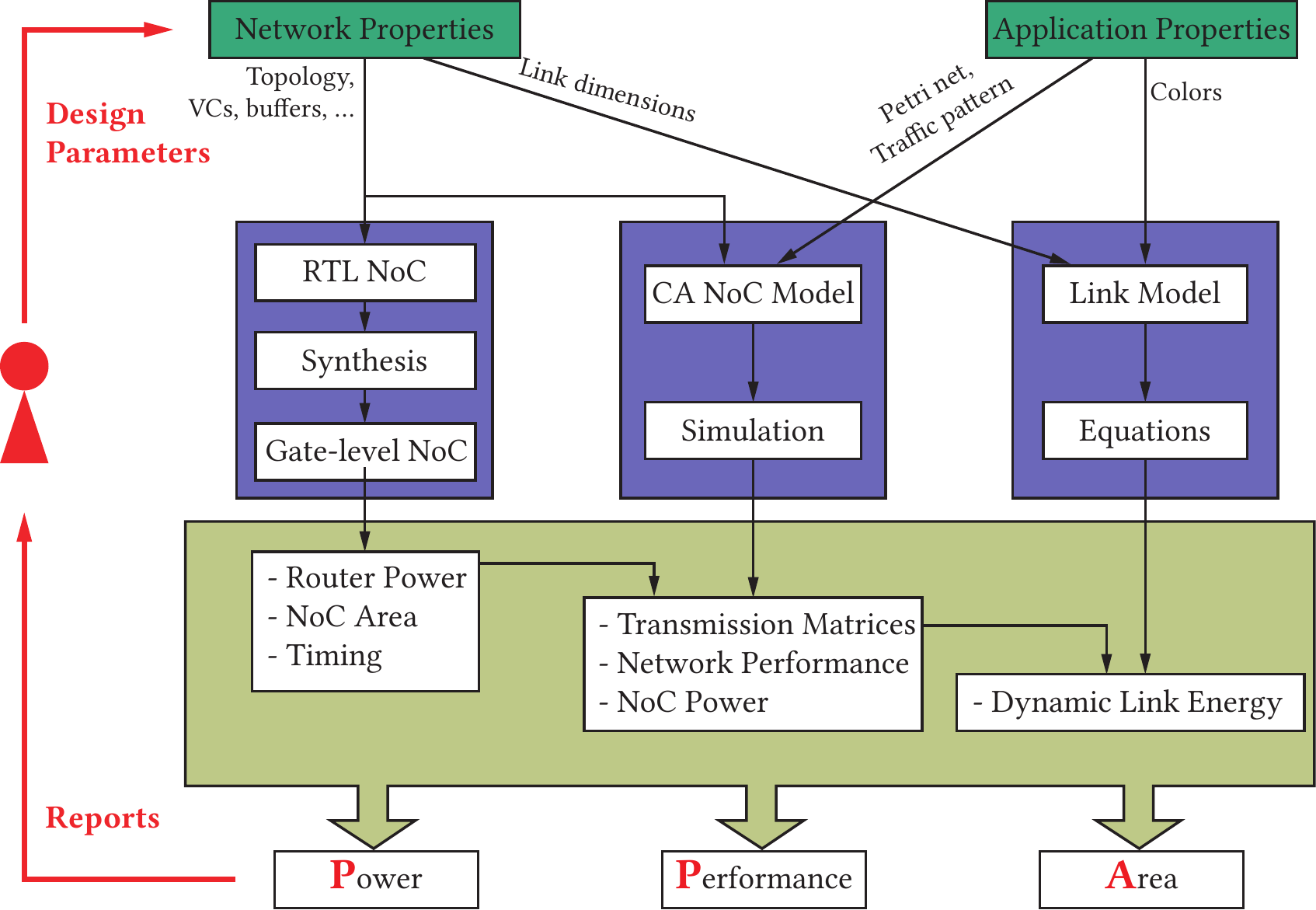}
	\caption{Tool flow}
	\label{fig:toolflow}
\end{figure}

The parts and functionality of \emph{Ratatoskr} is shown in Fig.~\ref{fig:toolflow}. Seen from the most abstract perspective, the tool flow of the framework is tailored towards in-depth PPA generation for the user: Design parameters are set and the framework generates PPA reports. Using these results, the user can modify the parameters until design constraints are met. 

There are two sets of \textbf{design parameters}, separated by their target: Network properties and application properties, as shown on top of Fig.~\ref{fig:toolflow} on top in green boxes. The network properties include the network topology and floorplan, the number of virtual channels (VCs), the buffer depths, used routing algorithm and the link dimensions; the application properties define either a synthetic traffic pattern or an instance of the application model (cf. Sec.~\ref{sec:models:app}). Setting the parameters is simple, as only a single-point of entry file \texttt{bin/config.ini} is modified and a Python script \texttt{bin/configure.py} is executed. They set up the whole framework. The detailed options for parameters are introduced in Sec.~\ref{sec:tool:config}. The Python script \texttt{bin/plot\_network.py} opens a GUI and displays the network with a floorplan. 

The actual execution of the framework starts after setting the design parameters using the aforementioned Python script. In general, there are three parts in \emph{Ratatoskr}, which reflect the different levels of abstract present. The parts are depicted in Fig.~\ref{fig:toolflow} in blue boxes; they are
\begin{itemize}
	\item \textbf{RTL NoC:} The box on the left-hand side shows the hardware model of the NoC:  On RT level, a NoC is generated using Python scripts for meta-programming of VHDL code from the network properties. The implemented router is the novel pseudo-mesochronous vertical high-throughput router as published in our previous work \cite{Joseph.2018b}. The properties of this router are discussed in detail in Sec.~\ref{sec:model:router}. 
	
	The advantage of an RTL NoC over a purely cycle-accurate model is that it is synthesizable to gate level and thus it can be verified with a precise gate-level simulation with backannotated delays. Second, the RT level NoC can be synthesized for standard cell libraries. This generates precise data for NoC power, area and clock speed. Plus, it is possible to get results for heterogeneous 3D integration with different technology nodes. As a comfortable area-saving feature, our scripts remove unused parts of the crossbar and the allocator automatically based on information about the routing algorithm. This allows increasing area efficiency by up to 30\% over conventional routers, as synthesis for a commercial digital technology demonstrates. 
	
	The whole RTL NoC can be simulated using VHDL simulators. We provide processing elements that inject uniform random traffic as well as a trace-file based traffic generator for real-world application traffic. While RTL simulation is very precise, it is also very slow and only a few thousand clock cycles can be simulated realistically. Therefore, we provide a higher simulation performance of the NoC on CA level.
	
	\item \textbf{CA NoC Model:} The aforementioned NoC simulator on CA level is shown on the blue box in the middle of Fig.~\ref{fig:toolflow}. It actually takes both the network and the application parameters as input. The latter are required to load realistic traffic patterns into the network. The CA router model copes with the non-purely synchronous transmission of data, which is typical for heterogeneous 3D systems. This is a novel feature not to be found in the competing simulators. The simulator can be run using the Python scripts \texttt{bin/run\_simulations.py} or by directly running the executable. Since the simulator is a complex software and the core of the framework, Sec.~\ref{sec:impl:simulator} is dedicated to the technical details.
	
	The simulator generates results for the network performance (e.g. flit latencies) and the dynamic router power. We use the power model of Noxim, in which power-relevant events are counted, cf. \cite[Sec. 5.1]{Fazzino.2008}). As an innovative feature, the simulator stores transmission matrices. In short, these report the transition probabilities between idle and non-idle states and the data types transmitted (i.e. modeled by the colors in the application model) of all links in the network. This feature allows for precise dynamic and pattern-dependent link energy, as explained in the next paragraph. 
	
	\item \textbf{Link model:}  To generate precise power results for the links, we use our power models for single (vertical and horizontal) links \cite{Bamberg.2017, Bamberg.2018} and integrate them into our framework. This requires using the aforementioned transmission matrices. It enables precision within 1\% of much slower bit-level simulations and also allows for post-simulation assessment of different hop-to-hop data codings without a second simulation execution. The power models, implemented in Python, can be found in the \texttt{power} folder.
\end{itemize}

After finishing the execution of the framework, PPA results are generated as shown in the bottom-most part of Fig.~\ref{fig:toolflow}. Beside the individual results of the framework's parts, which already have been described before, \emph{Ratatoskr} also encapsulates the most relevant ones into reports. This maintains high usability and is in-line with our approach to a rapid design space exploration by iterating design parameters and testing them against constraints.

\subsection{Setting design parameters} \label{sec:tool:config}

\begin{table}
	{\footnotesize
	\centering
	\begin{tabular} {|l|l|p{4.4cm}|p{5.5cm}| } 
		\hline
		\textbf{Property} & \textbf{Section} & \textbf{Field} & \textbf{Description} \\
		\hline\hline
		\textcolor{col2}{Application} & \textcolor{col1}{Config} & \multicolumn{2}{l|}{\emph{\textcolor{col1}{General parameters such as simulation time and benchmark.}}} \\ \cline{3-4} 
		\textcolor{col2}{Properties}	&&simulationTime& length of simulation in ns\\\cline{3-4}
			&&flitsPerPacket& maximum length of packets\\\cline{3-4}
			&& benchmark & ["synthetic", "task"] to select application model\\\cline{2-4}
		& \textcolor{col1}{Task} & \multicolumn{2}{l|}{\emph{\textcolor{col1}{Configuration of task (application)  model traffic.}}} \\\cline{3-4} 
			&& libDir & folder path for configuration data \\\cline{2-4}
		& \textcolor{col1}{Synthetic} & \multicolumn{2}{l|}{\emph{\textcolor{col1}{Configuration of synthetic uniform random traffic.}}} \\\cline{3-4} 
			&& simDir & folder path for temporary data \\\cline{3-4}
			&& restarts & number of simulations  \\\cline{3-4}
			&& warmupStart, warmupDuration,  warmupRate & length and injection rate in warmup phase  \\\cline{3-4}
			&& \makecell[l]{runRateMin,\\ runRateMax,\\ runRateStep} & injection rates for the simulation \\\cline{3-4}
			&& runStartAfterWarmup, runDuration & timing of main run phase \\\cline{3-4}
			&& numCores & number of cores used  \\\cline{2-4}	
		& \textcolor{col1}{Report} & \multicolumn{2}{l|}{\emph{\textcolor{col1}{Configuration of reports.}}} \\ \cline{3-4} 
			&& bufferReportRouters & list of routers' ids to be reported \\\cline{3-4}		
	\hline\hline
	\multicolumn{2}{|l|}{\textcolor{col2}{Network}} & \multicolumn{2}{l|}{\emph{\textcolor{col1}{Hardware configurations like Network and Router properties.}}}\\\cline{3-4}
			\multicolumn{2}{|l|}{} & z & layer count\\\cline{3-4}
			\multicolumn{2}{|l|}{\textcolor{col2}{Properties}} & x, y & list of router counts per layer\\\cline{3-4}
			\multicolumn{2}{|l|}{} & routing& routing algorithm \\\cline{3-4}
			\multicolumn{2}{|l|}{} & clockDelay & list of clock delays per layer \\\cline{3-4}
			\multicolumn{2}{|l|}{} & bufferDepthType & ["single", perVC"] single or VC-wise buffer depth \\\cline{3-4}
			\multicolumn{2}{|l|}{} & bufferDepth & value for buffer depth \\\cline{3-4}
			\multicolumn{2}{|l|}{} & buffersDepths & list of buffers depths (one for each VC)\\\cline{3-4}
			\multicolumn{2}{|l|}{} & vcCount & list of VCs per router port\\\cline{3-4}
			\multicolumn{2}{|l|}{} & topologyFile & file with network topology \\\cline{3-4}
			\multicolumn{2}{|l|}{} & flitSize & bit-width of flits \\\cline{3-4}
	\hline
	\end{tabular}}
	\caption{A description of software and hardware configurations using the configuration \texttt{ini} file.}
	\label{tab:config}
	\vspace{-26pt}
\end{table}

The design parameters that can be set using the \texttt{config.ini} file are shown in Table~\ref{tab:config}. As already introduced, network and application properties are configured separately. 

The \textcolor{col2}{application} configuration has four sections. In \textcolor{col1}{Config}, the general parameters of the simulation are given: The simulation time and the maximum length of packets are set for the network; the selection between synthetic or application model and a directory for application model files are set for the benchmark. In \textcolor{col1}{Task}, the folder holding the two files for the application model is given. These are \texttt{xml} descriptions of the task graph and of the mapping of tasks to processing elements (cf. Section~\ref{sec:models:app}).  In \textcolor{col1}{Synthetic}, synthetic traffic patterns are configured: The number of simulation runs, used CPU cores and the temporary directory are given. Synthetic patterns have multiple phases and the duration of warm-up and run phases can be set. Furthermore, in \textcolor{col1}{Report}, the reports can be configured. A list of routers can be defined that are included in the generation of statistics which allows excluding edge-cases such as routers at the borders of the network.

The \textcolor{col2}{network} configuration has no sections. Here, all parameters of NoC and routers are set. This includes the number of layers in a 3D chip as well as the router count per dimension for conventional 3D mesh topologies. Note that the dimensions are a list to implement floorplans with different router counts per layer as found in heterogeneous integration. If other topologies than mesh are desired, a configuration must be done via separate Python scripts. Plus, the clock delay is set by a list, because of varying clock frequencies in heterogeneous 3D systems. Furthermore, the VC count and buffer depths are configured; this is possible per router or per individual VC. Next, a file path to the network topology is given. Finally, we set the bit-width of flits. 

It is possible to configure the network and the application for the simulator in further extend by modifying the intermediate \texttt{xml}, which the configuration Python scripts generate as input for the next stage of the framework, especially the simulator. The \texttt{xml} files are described in Section~\ref{sec:impl:simulator}. 
	
\section{Architectures and models}\label{sec:models}

\subsection{Application model}\label{sec:models:app}

We start the description of the architecture and models on the top-most abstraction level: The application model on transaction level. It must be abstract yet accurate and at the same time account for all properties of typical application executed on system-level. Since we especially target heterogeneous 3D chips, the relevant application areas must be covered; This includes a broad spectrum of use cases: Many platforms have been proposed from high-performance processors \cite{Katti.2010} to Vision SoCs for image processing \cite{Zarandy.2011} and wireless sensors for IoT (Internet of Things) \cite{Garrou.2009}. Therefore, the application model must account for (1) the timing of processing, which may change depending on input data and implementation technology, (2) the dynamic effects of varying input data (i.e.\ the statistical expected behavior) and (3) the data types transmitted for precise power modeling (for pattern-dependent switching coupling activities). We discussed models for applications in heterogeneous 3D SoCs in \cite{Joseph.2018}. There, we argue in detail that \emph{colored statistical Petri nets with retention time on places} are able to model all effects required. The colors are used to differentiate varying data streams with respect to pattern-dependent coupling switching. 

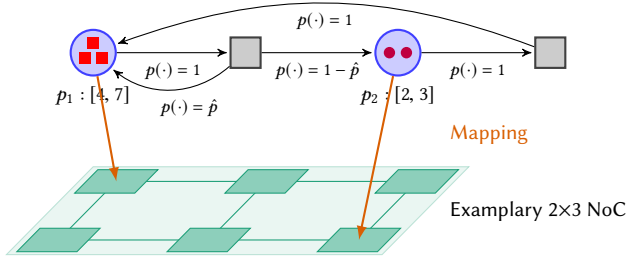
\begin{figure}
	\centering
	\input{fig/petrinetColoredTimedStochastic}
	\caption{Example for mapping of application model to NoC.}
	\label{fig:Petri}
\end{figure}

In general, a Petri net is an application model using graphs, in which data flows along edges are abstracted by tokens transmitted between places (vertexes). Our Petri nets model property (1) by retention times, which delay sending tokens; property (2) is modeled using a statistical net, in which sending tokens is associated with probabilities; property (3) is modeled by annotating tokens with colors that reflect the activities along the data flows. An example for such a Petri net is shown in Fig.~\ref{fig:Petri}. There are two places $p_1$ and $p_2$ modeling tasks in the application. The processing time on that places is in the given intervals: $[4,7]$ and $[2,3]$, respectively. With probability $\hat{p}$, the application sends tokens from $p_1$ to itself and with probability $1-\hat{p}$ to $p_2$. The tokens are colored, here depicted with red squares and purple circles (the shape is used for better accessibility).  In that very figure, we also show an exemplary layer of a NoC with 2$\times$3 mesh topology in  green. Each rectangle represents a tile comprising router, network interface (NI) and processing element (PE). The Petri-net is mapped on this network topology. The implementation of our application model and an introduction on its configuration is described in Section~\ref{sec:impl:simulator}.

\subsection{Router hardware architecture}\label{sec:model:router}

\begin{figure}
	\centering
	\includegraphics[width=10cm]{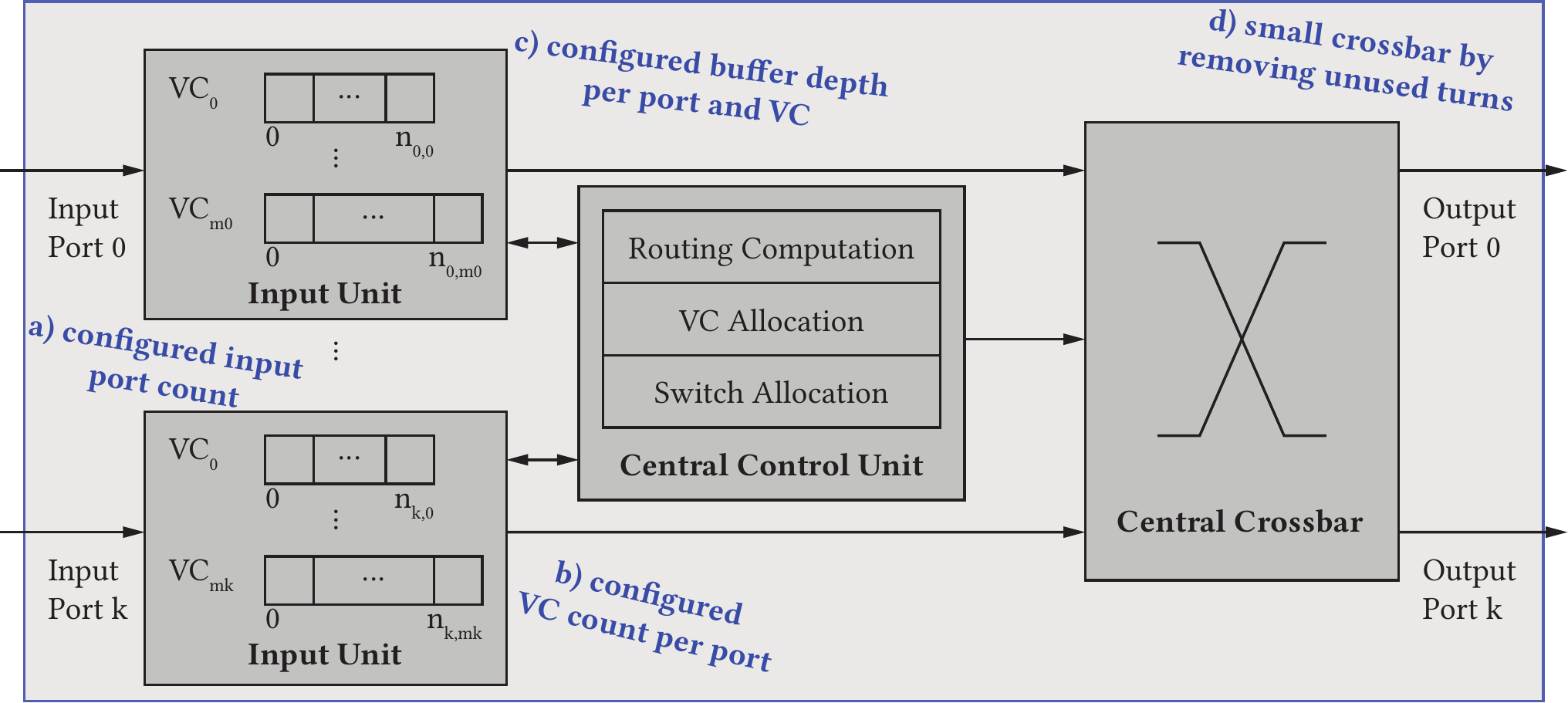}
	\caption{Router schematic.}
	\label{fig:schematic:router}
\end{figure}

We use a lightweight router architecture both on RT and CA level that suits the needs of heterogeneous 3D interconnects. By configuration interfaces, we also enable rapid prototyping. The technical details of the router are the following: We uses wormhole packet switching to reduce buffer depths, which are rather expensive especially in mixed-signal nodes. Flow-control is realized with credit counters. Number of ports is flexible in this design; therefore, it can be used in networks with different topologies. This is highly relevant for applications in heterogeneous 3D SoCs since irregular topologies are to be expected. Both simulator and router support varying flit widths. The router architecture is shown in Fig.~\ref{fig:schematic:router}. It is split into three major parts: In the \emph{input ports}, the flits are stored in buffers. The model implements VCs and the number of VCs and the buffer depth per VC can be set individually per router and per port. Thereby, smaller routers for instance in mixed-signal layers can be realized by e.g. by smaller buffers depths or less VCs. The \emph{central control unit} uses the architecture proposed by \cite{Becker.2009}, building the most light-weight router possible, which allows to use the router in expensive technology nodes as well (such as mixed-signal layers). Requests from the input ports will be processed and acknowledgments will be generated if a path is free. The central control unit has a modular design and consists of routing calculation, VC allocation, port allocation and switch allocation units. The VC arbiter prioritizes VCs with lower number, i.e.\ the first free VC with the lowest number is chosen. The allocation is done in input-first-manner; thus, one VC per input port requests and output port per clock cycle. The next VC may send a request after the previous VC received an acknowledgment, following round robin. The output port assigns acknowledgments also using round robin. The switch allocation does not perform maximum matching due to its large costs. Rather, a separable-input-first allocation is used for the switch allocation. It offers lower costs, but saturation rate is higher; this issue for saturation is actually negligible for the communication networks with only few VCs \cite{Becker.2009}. Routing decisions are made in routing computation unit. The \emph{crossbar} connects input ports and output ports. It has MUX-based architecture to minimize the control signals. We do not connect inputs and outputs, which are not possible based on the routing algorithm (cf. implementation details in Sec.~\ref{sec:impl:router}). Thereby, the size of the crossbar may be reduced by a significant portion, as explained in the results in Sec.~\ref{sec:results:router}. To summarize, our router supports rapid testing of different design by means simple setting of the following design parameters, as shown in Fig.~\ref{fig:schematic:router}: 
\begin{enumerate}[label=\alph*)]
	\item the input port count, 
	\item the VC count per port, 
	\item the buffer depth per port and VC and
	\item the turns forbidden by the routing algorithm for automatic area-reduction of the crossbar and the allocator.
\end{enumerate}

\subsection{Power model}\label{sec:models:power}

The power models comprise the dynamic energy of the router and of the links in the network. We take both into consideration, since related work showed that links cause between \unit[17]{\%} in the Teraflop router \cite{Kahng.2009} and up to \unit[53.9]{\%} in the NOSTRUM $8\!\times\!8$-NoC \cite[Table 1, p. 4]{Penolazzi.2006} of the overall energy consumption of the network. Extensive work on the power consumption of routers has already been conducted: ORION 3.0 \cite{Kahng.2015} provides the most detailed models for energy consumption of the individual router's components. The framework abstracts the power consumption for different technology nodes. ORION 3.0 is not directly linked to a NoC simulator. The Noxim simulator includes power models into NoC simulations, as introduced in \cite[Sec. 5.1, pp. 14-15]{Catania.2016}. In essence, Noxim counts power-relevant events in the router such as buffer write and buffer read, routing calculation etc. The values for each event are gathered from other sources (e.g.\ bit-level simulations or models such as ORION). Because of these extensive related works, we do use the same power models as Noxim within our simulator; the implementation in C++ is briefly introduced in Section~\ref{sec:impl:power}. We do not include a further description of the models here and kindly refer to \cite[Sec. 5.1, pp. 14-15]{Catania.2016}. 

There has been some research on the power consumption of links. While ORION includes a basic power model, as introduced in Eqs. 10 and 11 in Ref.~\cite{Kahng.2015}, the effects of pattern-dependency are not accounted for. However, we showed in \cite{Bamberg.2018} that this can lead up to 79.77\% modeling error. One option to leverage this error would be bit-level accurate link simulations; this, however, is too slow for a simulation of a complete NoC. Another option is to characterize data flows by their typical switching activities, modeled using the colors, such that flows with similar coupling switching properties are annotated with the same color in the application model. This can be used to estimate switching activities precisely. Therefore, we introduce the concept of \emph{data-flow matrices} $\textbf{M}$ for each link in NoC simulations. The entries of a data-flow matrix $\textbf{M}$ depend on the NoC hardware, i.e.\ VC count, arbitration, buffer depth, topology, and the application. For the latter, we use the colored Petri-net application model (cf. Section~\ref{sec:models:app}), in which each transmission between two tasks in an application is annotated with a color $\sigma_i$ from the set of all colors $\Sigma = \{\sigma_1, \dots, \sigma_n\}$.  A data-flow matrix therefore denotes the transitions between different colors on that link and also denotes, if a link was idle or used. For $n$ colors, each data-flow matrix $\textbf{M}$ has the size $[0,1]^{2n+3\times2n+3}$; It has a row and column for each color both as active (data of that color have been sent) or idle (last time the link was active, data of that color have been sent), head flits (active/idle) and an initial state until data were sent the first time via this link. In Ref.~\cite{Joseph.2018}, we introduced a toy example using a simulation of a router without VCs, which is passed by two data streams with different colors and different injection rates of 0.5625 flits/cycle and 0.1875 flits/cycle. This results in a transmission matrix such as:
\begin{equation*}\scriptsize
\begin{split}
\bordermatrix{
	& \text{idle}	& \text{(head, data)}  &\text{(head, idle)}& (\sigma_1, \text{data})   & (\sigma_1, \text{idle})  &(\sigma_2, \text{data})  &(\sigma_2,\text{idle})     \cr
	\text{idle} &0.010& 0.000 &-&-&-&-&- \cr
	\text{(head, data)}& -&-&0.019&0.042&-&0.014&- \cr
	\text{(head, idle)} &-&-&0.007&0.013&-&0.005&- \cr
	(\sigma_1, \text{data}) & -&0.041&-&0.326&0.126&-&- \cr
	(\sigma_1, \text{idle}) &-&0.013&-&0.113&0.039&-&- \cr
	(\sigma_2, \text{data})  &-&0.014&-&-&-&0.114&0.045\cr
	(\sigma_2, \text{idle})    &-&0.006&-&-&-&0.040&0.014\cr
}	
\end{split}
\end{equation*}
To highlight the expressiveness of the matrix, it shows this example that 32.6\% of all transmissions of color $\sigma_1$ are followed by the same color (Matrix read row-wise). Based on these pieces of information, the power consumption of the network can be calculated: Different colors represent data with different coupling and switching activities. Thus, the power consumption for packets with the same color is similar. This abstraction allows for precise and fast power estimation. At this point, we do not dig deeper into the details of the power model such as the physical formulae, as it has been published in Ref. \cite{Bamberg.2018}. However, the implementation of the power model in Python is discussed in Sec.~\ref{sec:impl:power}. The performance and accuracy of our approach are introduced and discussed in Sec.~\ref{sec:res:power}.

\section{Implementation}\label{sec:implementation}

\subsection{NoC simulator using C++ and SystemC}\label{sec:impl:simulator}

The NoC simulator inside the \emph{Ratatoskr} framework is implemented in C++14 using SystemC 2.3.3 class library, which provides a simulation kernel. All design parameters can be set without modifications to the code and therefore without recompiling. Only if novel functionality is required, the core will change; this is easy due to our clear class hierarchy. We start by explaining our software architecture. Next, we dig into implementation details of interesting components.


\subsubsection{Software architecture}

\begin{figure}
	\centering
	\includegraphics[width=.6\textwidth]{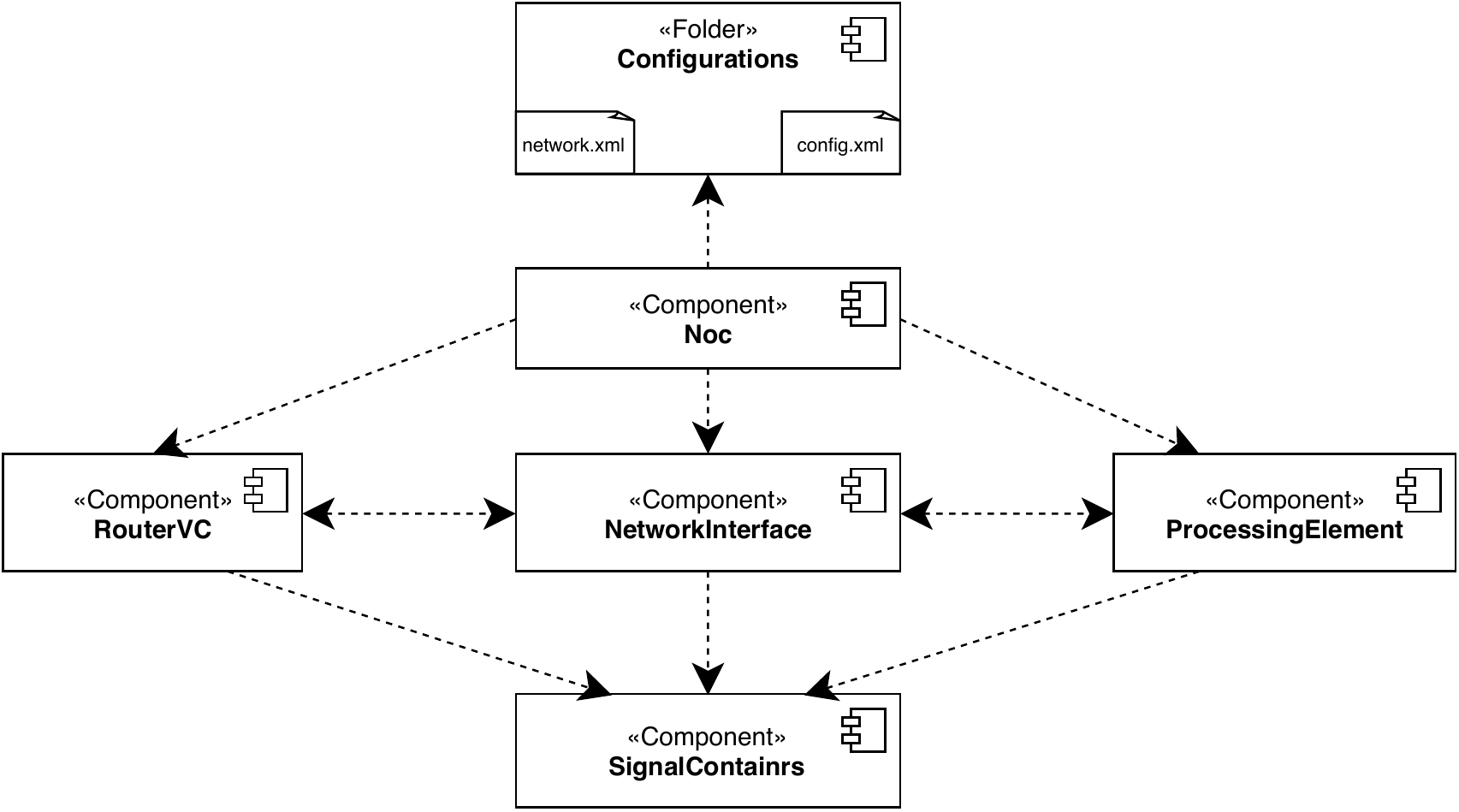}
	\caption{Component diagram of the main components.}
	\label{fig:compo_diag}
\end{figure}

A component diagram of the main parts is depicted in (Fig. \ref{fig:compo_diag}). The dotted arrows represent the dependency between two components (the component at the arrow's source depends on the component at the arrows' end). We did not draw all the components of the system for the sake of brevity. One can see that the \texttt{NoC} class reads the configuration of the simulation from the \texttt{config.xml} file and the configuration of the NoC from \texttt{network.xml}, both located in a configuration folder. The \texttt{NoC} class then hosts all components of the network: First, the \texttt{RouterVC} class implements the router model with VCs. Second, the \texttt{NetworkInterface} class interconnects PEs and routers. Third, the \texttt{ProcessingElement} class implements the PEs by providing a platform to simulate the application model and hosting tasks. All three components are connected by SystemC signals for data, flow control, etc. We abstract these in the \texttt{SingalContainer} class for easy use and maintenance.

The simulator inside \emph{Ratatoskr} provides two router models, a small standard router based on \cite{Becker.2009} and a vertical high-throughput link/router \cite{Joseph.2018b}, as well as four routing algorithms. Furthermore, it is possible to extend the code by adding new classes (e.g. novel router models or other routing algorithms) that implement the abstract classes provided by our solution (see next paragraphs and Fig.~\ref{fig:class_diag}). Therefore, one only has to write a C++ implementation, then register it in the constructor of \texttt{Router} base class and compile.

\subsubsection{Implementation details}
\begin{figure}
	\centering
	\includegraphics[width=0.9\textwidth]{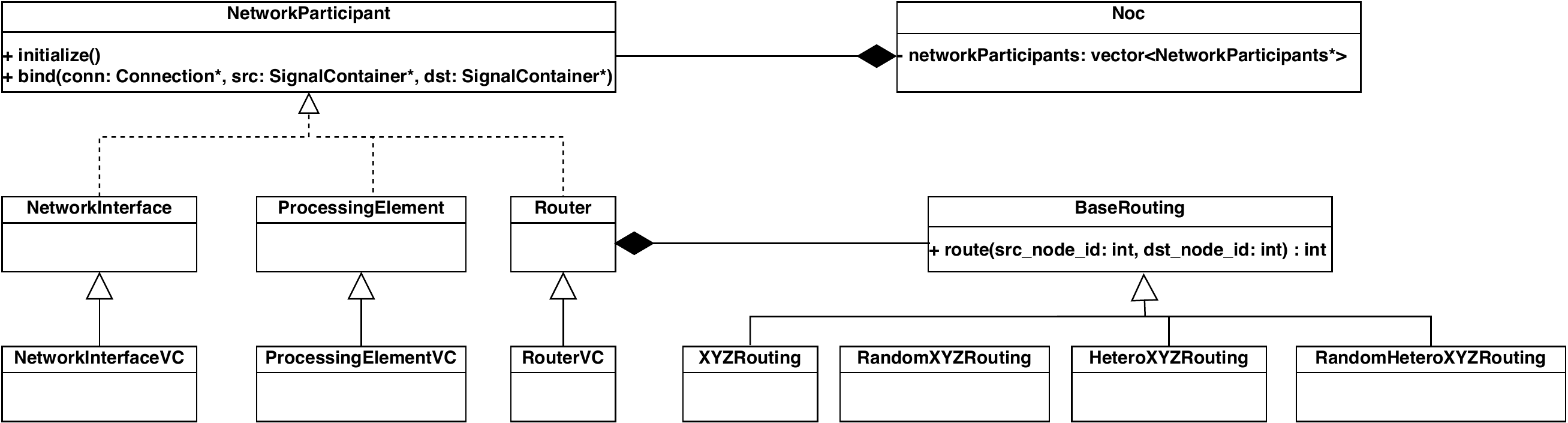}
	\caption{Class diagram that shows the modular design of our NoC simulator.}
	\label{fig:class_diag}
\end{figure}

In this section we discuss the implementation details and the benefits of such design.
In Fig.~\ref{fig:class_diag}, we see the general software architecture of the \emph{Ratatoskr} framework. The top-level class is \texttt{Noc}. All NoC components are inheriting form the \texttt{NetworkParticipant} class. The \texttt{NoC} class is the top-level class in the simulator. It contains a vector with all \texttt{NetworkParticipant}s. Each \texttt{NetworkParticipant} provides two virtual functions: \texttt{initialize} that initializes components and \texttt{bind} that binds the signals of connections between components. This directly follows the SystemC-modeling style. The NoC simulator provides three types of network participants: \texttt{Router}, \texttt{ProcessingElement} and \texttt{NetworkInterface}. Each one of these classes is realized through a concrete class implementing the virtual functions of their base classes (namely \texttt{initialize} and \texttt{bind}). Furthermore, the behavioral model of the components is implemented in the concrete classes with the suffix \texttt{VC}. We use modern C++ (C++14) extensively within these models; this does not only enable reader-friendly code but also a more abstract modeling. For instance, within the router there are no further SystemC sub-modules; rather, the communication within the router between VC allocation, arbitration and switch, as introduced in Section~\ref{sec:model:router}, is realized via data structures from the standard library. For instance, request and acknowledgments are \texttt{std::maps} that constitute the router architecture in data. The advantage lies in the option to iterate these data structures. This flat hierarchy is advantageous both for better software maintenance and higher simulation performance, as SystemC modules and communication via ports result in context switches during simulation. The implementation of routing functions is also shown in Fig.~\ref{fig:class_diag}, on the bottom-right part. There is a base class \texttt{BaseRouting}, which has a virtual function \texttt{route}; it takes the current node's address and the destination's address as input and calculates an output port. We provide a set of deterministic, low-overhead routing algorithms for heterogeneous 3D SoCs as published in \cite{Joseph.2018b}. To summarize, we used inheritance and polymorphism programming paradigms to achieve a highly maintainable and flexible architecture. The user can freely add implementations of new participants or routing algorithms without breaking or recompiling the code base.


\subsubsection{Detailed configuration using XML files}

Using \texttt{xml} files allows for a very fine-grained configuration of the NoC under test. We provide an example of the hardware description in Fig.~\ref{fig:networkxml}. The file starts with the definition of routers and PE using the \texttt{nodeTypes} tag. Nodes are identified using unique ids. First, we show the definition of a router with VCs using dimension-ordered routing and \unit[1]{GHz} clock frequency. Second, a PE is defined with VCs and \unit[1]{GHz} clock frequency. Next, the nodes are instantiated in the \texttt{nodes} tag. Each \texttt{node} has a position in 3D space, is associated with one of the \texttt{nodeTypes} and has a unique id. Finally, \texttt{connections} between nodes are defined. Each connection has two \texttt{ports}. The end-points of the ports are connected using the unique node ids; furthermore, the VC count is defined and the buffer depth is set (either for all VCs or VC-wise). Via the connection's data structure any topology for a heterogeneous 3D SoC can be defined. Since the \texttt{xml} files tend to be rather long, we provide Python classes for their functional definition.

We also provide an example for the application description using colored statistical Petri-nets with retention time on places in the code shown in Fig.~\ref{fig:appxml}. One exemplary task, which implements places, is defined there. Tasks have unique ids. In-going and out-going data connections are configured with the \texttt{requires} and \texttt{generates} tags. Application data are generated after all required data, from the \texttt{requires} tag, are available. To model the stochastic property, data can be sent to different \texttt{destinations}, which are grouped into a \texttt{possibility} each that is selected following a given \texttt{probability}. The timing of tasks and the retention time on places is modeled using the tags \texttt{start}, \texttt{duration}, \texttt{repeat}, \texttt{interval} and \texttt{delay}. A task is only executed in between the \texttt{start} time and until its \texttt{duration} is finished. A task will also stop execution, if it was repeated for as many times as given by the tag \texttt{repeat}. For each repetition, one data send possibility is taken. If tokens are available, a task sends data to this destination. It will send data \texttt{count} times. There is a delay in-between sending data, in which the task is idle; this implements the retention time. The colored Petri net is realized by defining data types as shown in the lower part of the code in Fig.~\ref{fig:appxml}. We do not provide an exemplary mapping file because of its simple structure mapping task ids to PE node ids.

\input{code/NoC}

\subsection{NoC router using VHDL}\label{sec:impl:router}

The implementation of the router in VHDL is straight-forward following the modular architecture introduced in Sec.~\ref{sec:model:router}. It can be configured for the aforementioned design parameters by using the python scripts \texttt{hardware/vhdl\_writer.py}. These take the \texttt{config.ini} as an input and write VHDL sources, accordingly. The script also generates a directory as input for synthesis, e.g.\ with Synopsis design compiler. There are four network options:
\begin{enumerate}
	\item A NoC using a conventional router as introduced in Sec.~\ref{sec:model:router}.
	\item A NoC using the high-throughput router as introduced in \cite{Joseph.2018b}.
	\item A NoC with PEs injecting uniform random traffic using the conventional router.
	\item A NoC with PEs injecting uniform random traffic using the high-throughput router.
\end{enumerate}
The two latter options can be used for verification as well as VHDL simulation. Also, the NoC can be synthesized for FPGA-based prototyping \cite{Drewes.2017}, as well. 

In addition to the configuration options, we also target low area overhead. Therefore, we remove turns from the crossbar that are impossible within the routing algorithm to reduce its size. We use a data structure, in which the possible and impossible turns are stored. Based hereupon, the crossbar either connects the input to the output port or both to ground. In the latter case, the synthesis tool optimizes the links and, automatically, the size of the crossbar is reduced without further user interaction.

We also provide a VHDL traffic generator and receiver based on trace files generated by Python scripts that are connected to our simulator (or any other high-level tool to generate traffic patterns). A so-called "Data Generate Unit" then injects the traffic based on the patterns information on the packet length and the injection time. Furthermore, we ship a "Data Converter Unit". It allows to convert the received data from the network to a end-user-friendly data format. Specifically, it allows to: (1) back/conversion of received data to the original data type and do further processing (like noise analysis); (2) comparison of the received data (from hardware simulation) and the generated data (from high-level simulation) for verification and error analysis; (3) reporting of system statistics just as generated from the high-level model. This hardware-level functionality gives the user an extended framework for verification, prototyping and emulation.

\subsection{Power models using Python}\label{sec:impl:power}

The dynamic router energy is implemented just as in Noxim. Five events are tracked: A buffer write, a buffer read, popping the head element from a buffer, routing calculation and crossbar traversal. The occurrences of these events are counted using a \texttt{power} class, which is singleton. 

The dynamic link energy is calculated using the aforementioned data-flow matrices during simulation, which then are fed into a Python implementation of the power model along with the correct parameters for link width and size, switch activities represented by colors in the application model and the used technology nodes. The data-flow matrices could be generated during data sending in the routers. Although this would slightly increase simulation performance, we implemented a separate \texttt{Link} class. This allows for better maintainability and readability of the code. The links are modeled cycle-accurate and add an entry to the correct entry in the data-flow matrices in each clock cycle. Links are modeled unidirectional, because of non-purely synchronous interaction between routers. When a simulation stops, the data-flow matrices are written into \texttt{csv} files that then are read by the aforementioned Python model. The Python scripts are given in the folder \texttt{power}. An \texttt{interconnect} package implements four classes: The class \texttt{Driver} implements the model for the driver of a link, and the class \texttt{Interconnect} implements the physical models for the link itself. The class \texttt{DataStream} and \texttt{DataStreamProb} implement the properties of data streams (i.e.\ colors in the application graph). An exemplary usage of the link power model is shown in Fig.~\ref{fig:linkexample}. First, the properties of a 3D link are defined. Next, the parameters are passed to the link model. As additional features, the link model does not only calculate power but can also be used for other physical properties of the links. For instance, one can obtain the maximum length (in mm) based on target clock frequency (and vice versa); the link area including keep-out-zones is also provided for any link design. Finally, three data streams are defined. The first is based on samples and the second and third on expected behavior. Finally, the energy is calculated using these data streams and a data-flow matrix as given from the simulator.

\begin{figure}
\begin{lstlisting}[style = linkmodel]
from interconnect import Interconnect, Driver, DataStream, DataStreamProb

# define 3D link
phit_width = 16  #transmitted bits per link (incl. flow-control, ECC, etc.)
wire_spacing, wire_width = 0.6e-6, 0.3e-6
TSV_pitch, TSV_radius, TSV_length = 8e-6, 2e-6, 50e-6 # length is constant
metal_layer = 5
ground_ring = False  # structure used to reduce TSV noise (affects power)
KOZ = 2 * TSV_pitch
driver_40nm_d4 = Driver.predefined('comm_45nm', 4)  # comm. 40nm driver / or define own via Driver()

# set parameters of link
interconnect_3D_dig = Interconnect(B=phit_width, wire_spacing=wire_spacing, wire_width=wire_width,
	metal_layer=metal_layer, Driver=driver_40nm_d4, TSVs=True, TSV_radius=TSV_radius,
	TSV_pitch=TSV_pitch, TSV_length=TSV_length, KOZ=KOZ, ground_ring=ground_ring)

# get maximum length based on target clock frequency
f_clk_dig = 1e9
l_max_3D_dig = interconnect_3D_ms.max_metal_wire_length(f_clk_dig)

# get area of link
TSV_area = interconnect_3D_dig.area_3D

# define three datastreams with different properties
ds1 = DataStream(ex_samp, B=16, is_signed=False)  # 16b data stream from specific samples
ds2 = DataStream.from_stoch(N=1000, B=16, uniform=1, ro=0.4)  # random dist. ro := correlation (1000 samples)
ds3 = DataStream.from_stoch(N=1000, B=16, uniform=0, ro=0.95, mu=2, log2_std=8)  # gaussian

# calculate energy based on data flow matrix from matrix
E_mean = interconnect_2D_dig.E([ds1, ds2, ds3], data_flow_matrix) 
\end{lstlisting}
\caption{Exemplary usage of the link power model.}\label{fig:linkexample}
\end{figure}

\section{Available PPA reports}\label{sec:reports}

Our proposed tool provides the PPA measurements for the NoC in the whole chip. In short, the following is provided:

\noindent\begin{minipage}[t]{0.35\textwidth}
	\textbf{Performance}\vspace{-2pt}
	\begin{itemize} [leftmargin=*] \setlength{\itemsep}{0pt}
		\setlength{\parskip}{0pt}
		\setlength{\parsep}{0pt}
		\item Mean and median flit delay 
		\item Mean and median  packet delay 
		\item Mean and median  network delay 
	\end{itemize}
\end{minipage}\hfill\begin{minipage}[t]{0.28\textwidth}
\textbf{Power}\vspace{-2pt}
\begin{itemize} [leftmargin=*] \setlength{\itemsep}{0pt}
	\setlength{\parskip}{0pt}
	\setlength{\parsep}{0pt}
			\item Dynamic router power (Noxim) 
			\item Dynamic link energy 
\end{itemize}
\end{minipage}\hfill\begin{minipage}[t]{0.28\textwidth}
\textbf{Area}\vspace{-2pt}
\begin{itemize} [leftmargin=*] \setlength{\itemsep}{0pt}
	\setlength{\parskip}{0pt}
	\setlength{\parsep}{0pt}
			\item Router area for a given technology node 
			\item Link area
\end{itemize}
\end{minipage}\vspace{4pt}

After simulation, the \emph{Ratatoskr} framework generates several reports about the aforementioned measurements. First, there is a textual report (\texttt{report.txt}); it is the most general and basic information, namely the average flit, packet and network latencies. Furthermore, the clock count and delay per layer and all (normalized) data-flow matrices per link are included. Second, \texttt{report\_Links.csv} contains flattened link data-flow matrices without normalization, for further automated processing. Third, \texttt{report\_Routers\_Power.csv} contains the dynamic power of each router. Forth, the usage of virtual channels is reported. The \texttt{VCUsage} folder contains VC usage per each router, with csv files named after router IDs. The rows of each file denote the ports of a router (in order: local, east, west, north, south, up and down) and the columns represents the count of VCs used, i.e.\ in the first row, all VCs are empty and in the last row all VCs are filled with at least one flit. Fifth and finally, the usage of buffers is reported, as well. The \texttt{BuffUsage} folder contains the buffer usage of the routers as \texttt{csv} files, named after router ID and direction. The rows of the file are equal to the buffer depth and the columns denote the VC numbers. Thereby, a usage count for each buffer element is given.

When executing the fully-automated tool flow using the Python scripts \texttt{run\_simulation.py}, \emph{Ratatoskr} collects all necessary pieces of information from the previous files and generates a pdf file with a visual summary. It includes three types of plots: The network performance, i.e.\ the latency over injection rate; the average buffer usage per layer and direction as 3D histograms and the average VC usage per layer as histogram. The exemplary plots are given in Fig.~\ref{fig:examplary_reports}. In Fig.~\ref{fig:latencies} the network performance is given for different injection rates for an exemplary network configuration, namely a 4$\times$4$\times$4-NoC with three heterogeneous technologies.\footnote{In this example, we use \unit[130]{nm}, \unit[65]{nm} and \unit[32]{nm} technology nodes.} Our reports include the standard deviation as well as the mean of the relevant latencies. In Fig.~\ref{fig:VC_Usage} the VC usage is given for the lower layer (therefore, the downwards direction is never used). One can see, that higher VC numbers are less used, which is in line with the router model. Also the pressure on west and east is higher, which is a consequence of XYZ routing and round-robin VC arbitration. Finally, in Fig.~\ref{fig:Buff_Usage} a histogram is given, which reports for one exemplary direction in a layer the number of times, a certain buffer usage and VC usage were given. These three reports available from \emph{Ratatoskr} give in-depth insight into the network dynamics and allow for router parameter optimization even on a micro-architectural level, e.g. for single buffer elements.

\pgfplotstableread[col sep = comma]{data/VCusage.csv}\VCusage
\begin{figure}
	\centering
	\begin{subfigure}[t]{.43\textwidth}
		\includegraphics[width=1.1\linewidth]{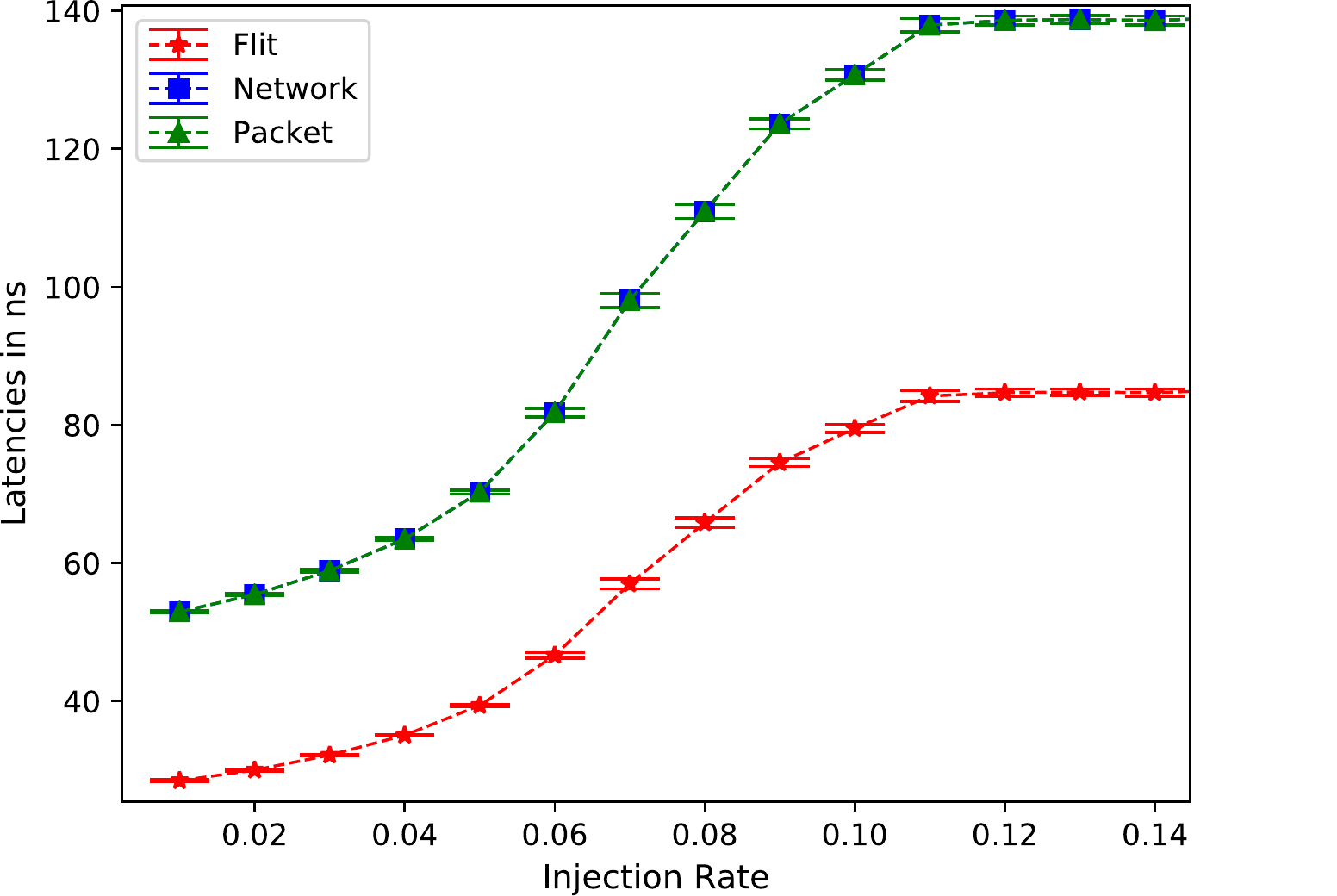}
		\caption{Network performance.}
		\label{fig:latencies}
	\end{subfigure}
	\hspace{1cm}
	\centering
	\begin{subfigure}[t]{.45\textwidth}
				\begin{tikzpicture}
				\begin{axis}[ybar,
				xmin=0.5,xmax=5.5,
				ymin=0,
				height=145,
				width=1.12\columnwidth,
				legend pos=north west,bar width=1,axis on top,style = {font=\scriptsize},
				xlabel={Number of used VCs},
				ylabel={Count},
				title = {Layer 0, Injection Rate = 0.07},
				xlabel style = {font=\scriptsize, yshift=1ex},ylabel style = {font=\scriptsize, yshift=-5ex},
				yticklabel style = {font=\scriptsize,xshift=.5ex},
				xtick=data,
				xticklabels from table={\VCusage}{vcusage},
				legend pos=north east,
				]
				\addplot [color=col1, fill = col1] table[x=vcusage,y=local] {\VCusage};
				\addplot [color=col2, fill = col2] table[x=vcusage,y=east] {\VCusage};
				\addplot [color=col3, fill =  col3] table[x=vcusage,y=west] {\VCusage};
				\addplot [color=red, fill = red] table[x=vcusage,y=north] {\VCusage};
				\addplot [color=blue, fill = blue] table[x=vcusage,y=south] {\VCusage};
				\addplot [color=green, fill = green] table[x=vcusage,y=up] {\VCusage};
				\legend{Local, East, West, North, South, Up};
				\end{axis}
				\end{tikzpicture}

		\caption{VC usage histogram.}
		\label{fig:VC_Usage}
	\end{subfigure}
	\\[6pt]
	\begin{subfigure}[t]{.36\textwidth}
		\centering
		\includegraphics[width=\linewidth]{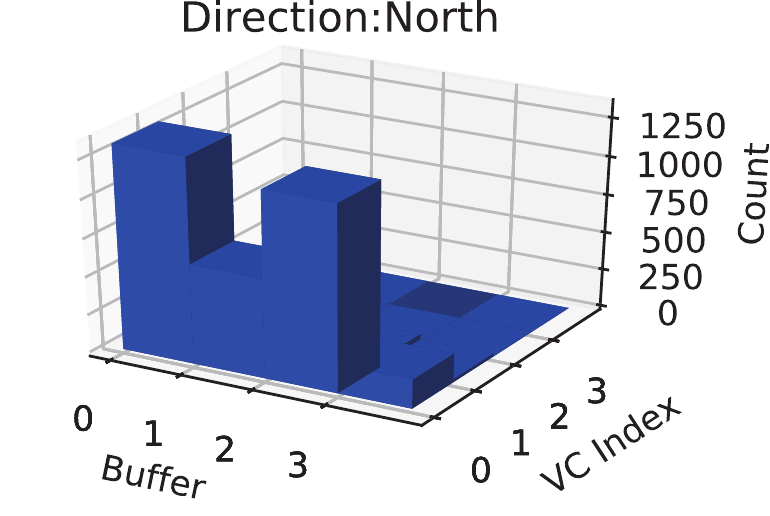}
		\caption{Buffer usage.}
		\label{fig:Buff_Usage}	
	\end{subfigure}
	\caption{Exemplary reports generated by \emph{Ratatoskr} after a simulation with uniform random traffic.}
	\label{fig:examplary_reports}
\end{figure}

To summarize, the automatically generated reports available by our framework provide in-depth insight into the NoCs PPA. Especially the buffer usage and the VC usage that extend the related work, as this feature is new. It is tailored towards heterogeneous 3D integration, because we average the reports per layer. Thereby, routers can be optimized in each layer, an thus per technology node, to meet both the technological and application's requirements.

%% file: fig/petrinetColoredTimedStochastic.tex
\begin{tikzpicture}[node distance=1.3cm,>=stealth',bend angle=45,auto]

\tikzstyle{place}=[circle,thick,draw=blue!75,fill=blue!20,minimum size=6mm]
\tikzstyle{red place}=[place,draw=red!75,fill=red!20]
\tikzstyle{transition}=[rectangle,thick,draw=black!75,
fill=black!20,minimum size=4mm]
\tikzstyle{tansitionlabel}=[font=\tiny]

\tikzstyle{every label}=[font=\scriptsize]

\node[place, label=below: {$p_1: [4, 7]$}](w1) {}
[children are tokens]
child {node [token, fill = red, shape = rectangle] {}}
child {node [token,fill=red, shape = rectangle] {}}
child {node [token,fill=red, shape = rectangle] {}};

\node[transition](t1) [right = 1.5cm of w1]{}
	edge [pre] node[tansitionlabel]{$p(\cdot)=1$} (w1)
	edge [post, bend left=40]node[tansitionlabel, xshift = .25cm]{$p(\cdot)=\hat{p}$} (w1)
;
\node[place](w2)[right = 1.5cm of t1, label=below:{$p_2: [2, 3]$}]{}
	[children are tokens]
	child {node [token, fill = purple] {}}
	child {node [token,fill=purple] {}}
	edge [pre]node[tansitionlabel]{$p(\cdot)=1-\hat{p}$} (t1)
;
\node[transition](t2) [right = 1.5cm of w2]{}
	edge [pre] node[tansitionlabel]{$p(\cdot)=1$} (w2)
	edge [post, bend right= 20] node[tansitionlabel]{$p(\cdot)=1$}(w1)
;

\tikzmath{\l1 = -1.5;};
\draw [col1, fill = col1!30, opacity= 0.3] (0,\l1, 0) -- (5,\l1,0) --(5,\l1,3) -- (0,\l1,3) -- cycle;
\draw[col1] (0.5, \l1, 0.5) -- (4.5, \l1, 0.5);
\draw[col1] (0.5, \l1, 2.5) -- (4.5, \l1, 2.5);
\draw[col1] (0.5, \l1, 0.5) -- (0.5, \l1, 2.5);
\draw[col1] (2.5, \l1, 0.5) -- (2.5, \l1, 2.5);
\draw[col1] (4.5, \l1, 0.5) -- (4.5, \l1, 2.5);
\draw [col1, fill = col1!50] (0.1, \l1, 0.1) -- ++(.8,0,0) -- ++(0,0,.8) -- ++ (-0.8,0,0) -- cycle;
\draw [col1, fill = col1!50] (2.1, \l1, 0.1) -- ++(.8,0,0) -- ++(0,0,.8) -- ++ (-0.8,0,0) -- cycle;
\draw [col1, fill = col1!50] (4.1, \l1, 0.1) -- ++(.8,0,0) -- ++(0,0,.8) -- ++ (-0.8,0,0) -- cycle;
\draw [col1, fill = col1!50] (0.1, \l1, 2.1) -- ++(.8,0,0) -- ++(0,0,.8) -- ++ (-0.8,0,0) -- cycle;
\draw [col1, fill = col1!50] (2.1, \l1, 2.1) -- ++(.8,0,0) -- ++(0,0,.8) -- ++ (-0.8,0,0) -- cycle;
\draw [col1, fill = col1!50] (4.1, \l1, 2.1) -- ++(.8,0,0) -- ++(0,0,.8) -- ++ (-0.8,0,0) -- cycle;
\node [desc, anchor = west] (source) at (5.2,\l1,1.5) {Examplary 2$\times$3 NoC};
\draw[col2, thick, -latex] (w1) -- (0.5, \l1, 0.5);
\draw[col2, thick, -latex] (w2) -- (4.5, \l1, 2.5);
\node [desc, anchor = west, col2] (mapping) at (5.2,-.5,1.5) {Mapping};

\end{tikzpicture}

%% file: code/NoC.tex
\begin{figure}
\begin{minipage}[b]{0.49\textwidth}
\begin{lstlisting}[style = myxml]
<nodeTypes>
	<nodeType id="0">
		<model value="RouterVC" />
		<routing value="XYZ" />
		<clockDelay value="1" />
	</nodeType>
	<nodeType id="1">
		<model value="ProcessingElementVC" />
		<clockDelay value="1" />
	</nodeType>
</nodeTypes>
<nodes>
	<node id="0">
		<xPos value="0"/>
		<yPos value="0"/>
		<zPos value="0"/>
		<nodeType value="0"/>
		<idType value="0"/>
	</node>
...
</nodes>
<connections>
	<con id="0">
		<ports>
			<port id="0">
				<node value="0"/>
				<bufferDepth value="16"/>
				<vcCount value="3"/>
			</port>
			<port id="1">
				<node value="8"/>
				<buffersDepths value="10, 20, 30"/>
				<vcCount value="3"/>
			</port>
		</ports>
	</con>
...
</connections>
\end{lstlisting}
\caption{Example for configuration of a NoC.}
\label{fig:networkxml}
\end{minipage}
\hfill
\begin{minipage}[b]{0.49\textwidth}
\begin{lstlisting}[style = myxml]
<task id = "1">
	<start min = "0" max = "0"/>
	<duration min = "100" max = "100"/>
	<repeat min = "2" max = "2"/>
	<requires>
		<requirement id = "0">
			<type value = "1"/>
			<source value = "0"/>
			<count min = "1" max = "1"/>
		</requirement>
		...
	</requires>
	<generates>
		<possibility id = "0">
			<probability value = "1"/>
			<destinations>
				<destination id = "0">
					<delay min = "0" max = "50"/>
					<interval min = "10" max = "10"/>
					<count min = "3" max = "3"/>
					<type value = "1"/>
					<task value = "3"/>
				</destination>
				...
			</destinations>
		</possibility>
		...
	</generates>
</task>

<data>
	<dataTypes>
		<dataType id = "0">
			<name value = "image"/>
		</dataType>
		...
	</dataTypes>
</data>
\end{lstlisting}
\caption{Example for a configuration of an application.}
\label{fig:appxml}
\end{minipage}

\end{figure}

%% file: chapters/04_results.tex
\section{Results}\label{sec:results}


\subsection{Simulation performance}\label{sec:res:sim}

The \emph{simulation time for different injection rates} is shown in Fig.~\ref{fig:performanceInjRate}. We simulate a 4$\times$4 NoC with 32 flits per packet, 4 flit deep buffer, 4 VCs and dimension order routing (XY-routing) with Booksim 2.0 (depicted as green rectangles), Noxim (orange circle) and  our simulator inside \emph{Ratatoskr} (blue cross). Uniform random traffic is injected into the network. The injection rate is increased from 0.015 flits/cycle to 0.08 flits/cycle in steps of 0.05 flits/cycle. We run 10 simulations using Ubuntu 18.04 on a single core of an Intel i7-6700 at 4GHz. Fig.~\ref{fig:performanceInjRate} reports the median and the standard deviation of the run time. We did not run any other programs alongside (beside system services) to reduce side-effects. Both Noxim and \emph{Ratatoskr} do not change their performance with larger injection rates, while Booksim 2.0 does. For low injection rates, Booksim 2.0 is faster than the other two competitors; this relation is reversed for injection rates higher than 0.035 flits/cycle. Ratatoskr is consistently slower than Noxim with approximately four to eight seconds.

The \emph{simulation time for different network sizes} is shown in Fig.~\ref{fig:sizeperf}. We simulate a NoC of varying size with the same properties as in the last example. Uniform random traffic is injected into the network with an injection of 0.03 flits/cycle. The network size is increased from 4$\times$4 to 10$\times$10 in steps of 1. Again, we run 10 simulations on the aforementioned machine. Fig.~\ref{fig:sizeperf} reports the median and the standard deviation of the run time. The performance relation between the programs remains the same for all network sizes: Booksim is slower than \emph{Ratatoskr}, which is slower than Noxim.

\pgfplotstableread[col sep = comma]{data/performance/performance.csv}\Performance
\pgfplotstableread[col sep = comma]{data/size03/performance.csv}\SizeDrei
\begin{figure}
	\begin{minipage}{0.49\textwidth}
		\centering
		\begin{tikzpicture}
		\begin{axis}[
		xmin=0.012,xmax=0.083,
		ymin=0,
		height=135,
		width=1.12\columnwidth,
		legend pos=north west,bar width=1,axis on top,style = {font=\scriptsize},
		xlabel={injection rate [flits/cycle]},
		ylabel={median simulation time in [s]},
		xlabel style = {font=\scriptsize, yshift=1ex},ylabel style = {font=\scriptsize, yshift=-5ex},
		yticklabel style = {font=\scriptsize,xshift=.5ex},
		xticklabel style = {font=\scriptsize,yshift=-0.2ex,/pgf/number format/fixed,/pgf/number format/precision=5},
		scaled x ticks=false,
		error bars/.cd,
		]
		\addplot [densely dotted, mark options={solid}, color=col1, mark = triangle,mark size=3pt] plot [error bars/.cd, y dir = both, y explicit, error bar style={solid}]	
		 table[x=inj,y=booksim, y error = booksimstd] {\Performance};
		\addplot [color=col2, densely dotted, mark options={solid},mark=o,mark size=3pt]  plot [error bars/.cd, y dir = both, y explicit, error bar style={solid}] table[x=inj,y=noxim, y error = noximstd] {\Performance};
		\addplot [color=col3, densely dotted, mark options={solid},mark=x,mark size=3pt] plot [error bars/.cd, y dir = both, y explicit,  error bar style={solid}] table[x=inj,y=rata, y error = ratastd ] {\Performance};
		\legend{Booksim 2.0, Noxim, Ratatoskr}
		\draw[latex-latex](axis cs:0.075,4.11)--(axis cs:0.075,8.256) node [midway, anchor = east] {2.0$\times$};
		\draw[latex-latex](axis cs:0.08,8.545)--(axis cs:0.08,19.725) node [midway, anchor = east] {2.3$\times$};
		\end{axis}
		\end{tikzpicture}
		\caption[Performance injection rate relationship]{Relation between simulation performance and injection rate.}
		\label{fig:performanceInjRate}
	\end{minipage} \hfill
	\begin{minipage}{0.49\textwidth}
				\begin{tikzpicture}
			\begin{axis}[
			xmin=3.8,xmax=10.2,
			height=135,
			width=1.12\linewidth,
			xtick=data,
			legend pos=north west,bar width=1,axis on top,style = {font=\scriptsize},
			xlabel={NoC size in $n\!\times\!n$ mesh},
			ylabel={median simulation time in [s]},
			xlabel style = {font=\scriptsize, yshift=1ex},ylabel style = {font=\scriptsize, yshift=-4.7ex},
			yticklabel style = {font=\scriptsize,xshift=.5ex},
			xticklabel style = {font=\scriptsize,yshift=-0.2ex,/pgf/number format/fixed,/pgf/number format/precision=5},
			scaled x ticks=false,
			error bars/.cd,
			]
			\addplot [color=col1, densely dotted, mark options={solid},mark = triangle,mark size=3pt] plot [error bars/.cd, y dir = both, y explicit, error bar style={solid}]	
			table[x=size,y=booksim, y error = booksimstd] {\SizeDrei};
			\addplot [color=col2, densely dotted, mark options={solid},mark=o,mark size=3pt]  plot [error bars/.cd, y dir = both, y explicit, error bar style={solid}] table[x=size,y=noxim, y error = noximstd] {\SizeDrei};
			\addplot [color=col3, densely dotted, mark options={solid},mark=x,mark size=3pt] plot [error bars/.cd, y dir = both, y explicit, error bar style={solid}] table[x=size,y=rata, y error = ratastd ] {\SizeDrei};
			\legend{Booksim 2.0, Noxim, Ratatoskr}
			\end{axis}
			\end{tikzpicture}
			\caption[Performance NoC size relationship]{Relation between simulation performance and network size.}
			\label{fig:sizeperf}
	\end{minipage}
\end{figure}

\subsection{PPA of router hardware}\label{sec:results:router}

\pgfplotstableread[col sep = comma]{data/router/urandPerformance/results.csv}\PerformanceRouter
\begin{figure}
\centering		
		\begin{tikzpicture}
		\begin{axis}[
		xmin=0.008,xmax=0.081,
		height=100,
		width=\linewidth,
		legend pos=north west,bar width=1,axis on top,style = {font=\scriptsize},
		xlabel={injection rate [flits/cycle]},
		ylabel={latency [ns]},
		xlabel style = {font=\scriptsize, yshift=1ex},ylabel style = {font=\scriptsize, yshift=-4.7ex},
		yticklabel style = {font=\scriptsize,xshift=.5ex},
		xticklabel style = {font=\scriptsize,yshift=-0.2ex,/pgf/number format/fixed,/pgf/number format/precision=5},
		scaled x ticks=false,
		error bars/.cd,
		]
		\addplot [color=col1,densely dotted, mark options={solid},mark = triangle,mark size=3pt] plot [error bars/.cd, y dir = both, y explicit, error bar style={solid}]	
		table[x=inj,y=flitLat, y error = flitLatStd] {\PerformanceRouter};
		\addplot [color=col2, densely dotted, mark options={solid},mark=o,mark size=3pt]  plot [error bars/.cd, y dir = both, y explicit, error bar style={solid}] table[x=inj,y=packetLat, y error = packetLatStd] {\PerformanceRouter};
		\legend{Flit latency,Packet latency}
		\end{axis}
		\end{tikzpicture}
		\caption{Average flit and packet latency for different injection rates.}
		\label{fig:routerPerf}
\end{figure}
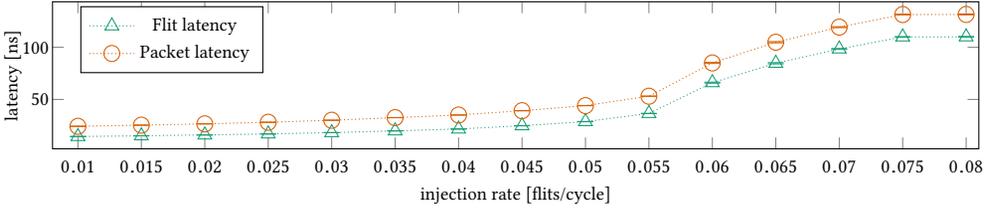

We ship an automatically configured and synthesizable hardware implementation along with each simulation run. Here, we shortly showcase the PPA figures of the high-throughput router architecture:

The network performance is shown in Fig.~\ref{fig:routerPerf}: The average flit latency and packet latency for different injection rates from 1\% to 8\% in [flits/cycle] are shown using a $4\!\times\!4\!\times\!4$ NoC with dimension-ordered routing, 4 VCs, 4-flit deep buffer and \unit[1]{GHz} clock speed. We simulated 100,000 clock cycles injecting uniform random traffic pattern. Both the average and the standard deviation are reported in the figure. 

To report area and power, we synthesize the same router for \unit[45]{nm} technology at \unit[250]{MHz} frequency. We compare three experimental setups, as shown in Tab.~\ref{tab:impossible_turns}: First, we use a router with a fully connected crossbar as baseline. Next, we use conventional XYZ dimension-ordered routing, which removes turns from the crossbar. This router does not account for the possible and impossible turns from the routing algorithm. Finally, we use Z\textsuperscript{+}(XY)Z\textsuperscript{-} routing algorithm from \cite{Joseph.2018b}. This routing algorithm shows very high performance for heterogeneous 3D SoCs and has less turns possible than conventional XYZ. We report the numbers for power in [\textmu W] and area in [\textmu m\textsuperscript{2}] for both the complete router and its crossbar.  One individual inner router has a total cell area of 37899 to \unit[39168]{$\mu m^2$} and a total power of 4.57e+03 to \unit[5.4e+03]{mW} depending on the routing algorithm. Table~\ref{tab:impossible_turns} also shows the area reductions possible from using information about the routing algorithms to reduce the size of the router crossbar. 
\begin{table}[h!]
	{\footnotesize\centering
	\begin{tabular}{l|r|r||r|r|r|r||r|r|r|r}
		\toprule
		& \multicolumn{2}{c||}{\textbf{Fully Connected}} &  \multicolumn{4}{c||}{\textbf{XYZ Routing Algorithm}} & \multicolumn{4}{c}{\textbf{Z\textsuperscript{+}(XY)Z\textsuperscript{-} Routing Algorithm}}           \\ \cline{2-11} 
		&                        Router            & Crossbar         & Router   & $\Delta$ & Crossbar & $\Delta$         & Router         & $\Delta$   & Crossbar  & $\Delta$         \\ 
		\midrule
		\textbf{Power} [\unit{$\mu W$}]                    & 5.40e+03          & 183.588          & 4.49e+03  &\textbf{-17\%}        & 64.005         &\textbf{-65\%} & 4.57e+03 &\textbf{-15\%} & 71.162   &\textbf{-61\%}          \\ \midrule
		\textbf{Area} [\unit{$\mu m^2$}]                     & 39168             & 1288             & 37942 &\textbf{-3\%}            & 894   &\textbf{-31\%}   & 37899    &\textbf{-3\%} & 880            &\textbf{-32\%}          \\ \bottomrule
	\end{tabular}}
\caption{Router cost reduction by removing impossible turns of routing algorithm}
\label{tab:impossible_turns}
\end{table}

\subsection{Power modeling capabilities}\label{sec:res:power}

To demonstrate the accuracy of our power models, we compare the estimated link power for a NoC with and without 4 VCs against a bit-level accurate simulation. The results are based on the case study provided in \cite[Sec. 8 and Fig. 7]{Joseph.2018c}, in which a 3D Vision SoC is simulated that consists of one layer in mixed-signal technology and one in memory technology. Six analog-digital converters in the mixed-signal layer send their 512$\times$512-pixel image data to a single memory in the adjacent layer via a NoC. The injection rate of image traffic is set to 20\% per sensor. As shown in Tab.~\ref{tab:power}, the used models are always within 2.4\% of the bit-accurate simulations, while conventional models that do not account for pattern-dependency and the impact of virtual channels yield an error of up to 42.5\%. For a NoC without VCs, both the conventional and \emph{Ratatoskr}'s power models yield a low error of \textless 1\%. Neither Booksim nor Noxim provide this power analysis feature; the accuracy for links is higher than ORION 3.0's power estimation (which is not embedded into simulations).

\begin{table}
	{\footnotesize
		\centering
		\begin{tabular} {l|lr|lr} 
			\toprule
			&\multicolumn{4}{c}{\textbf{Energy per transmitted flit [pJ]}} \\
			& without VCs & $\Delta$  & with four VCs&$\Delta$ \\
			\midrule
			\textbf{Baseline:} bit-accurate simulations & 2.39& & 4.18&\\
			Conventional model (without switching \& coupling) &2.40 & \textless 1\% &2.40 & 42.5\%\\
			Used model \cite{Joseph.2018c} &2.40 & \textbf{\textless 1\%}&4.28 & \textbf{2.4\%}\\
			\bottomrule
		\end{tabular}
	}
	\caption{Accuracy of used power models, numbers based on case study from Ref.~\cite{Joseph.2018c}.}
	\label{tab:power}
\end{table}

\section{Discussion}\label{sec:discussion}

The \textbf{simulation performance} is evaluated for varying injection rates and network sizes: First, concerning the \emph{simulation time for different injection rates}, see Fig.~\ref{fig:performanceInjRate}. Booksim 2.0 has a very high performance for small injection rates (by disabling router without traffic). However, the simulation speed is reduced linearly with higher injection rates. For injection rates higher than 0.02 flits per cycle, Noxim is faster than Booksim. For injection rates higher than 0.035 flits per cycle, \emph{Ratatoskr}'s simulator is faster than Booksim. In fact, for the highest injection rate evaluated (0.08 flits/cycle) Booksim 2.0 is 2.3$\times$ slower that \emph{Ratatoskr}. Both Noxim and \emph{Ratatoskr} have a constant simulation time independent from the injection rate, but \emph{Ratatoskr} is 2$\times$ slower than Noxim. This is not a result of less performing router or application model. The impact of the application model is very small, as we already evaluated in \cite[p. 6]{Joseph.2014}, where it was 1/30 of the overall simulation time. Rather, the slower performance is a direct result of the added functionality of the power model for links and detailed buffer usage statistics as for every link in every clock cycle, large data structures are written. This can easily be shown using profiling (using \texttt{gcc -pg}). Thus, our simulator is  slower than state-of-the-art but offers more features, but this is an acceptable compromise for the power model accuracy (see below). Second, concerning \emph{simulation time for different network sizes}, shown in Figure~\ref{fig:sizeperf}, one can see that all three simulators are linearly slower for more routers. Furthermore, the performance difference is constant. This is expected since every router is simulated for each clock cycle. To summarize, our simulator has the same performance as state-of-the-art if one does not account for extended features; the new features reduce performance but allow for better quality of results, as we will discuss in the next paragraph.

The \textbf{PPA of the router} is evaluated by synthesis of the high-throughput router for \unit[45]{nm} node. The results are shown in Table \ref{tab:impossible_turns}. We report the reduction in router power and area by removing the impossible turns of Z\textsuperscript{+}(XY)Z\textsuperscript{-} routing algorithm \cite{Joseph.2018b} and conventional XYZ routing algorithms. The power of the crossbar is reduced by 61\% and 65\%, respectively. This has a 15\% to 17\% positive effect of the total router power. The area of the crossbar is reduced by approx. 31\%--32\%; this reduces the total router area by approximately 3\%. The power and area enhancements are not affecting the router's performance, as the turns in the removed turns in the crossbar had not been taken. 

Our used \textbf{power models} are more accurate than conventional models without considering pattern-dependent switching coupling and virtual channels as clearly shown in Tab.~\ref{tab:power}: The error of our models is 2.4\% while conventional models yield up to 42.5\% error for a given case study. These very good results are a consequence of the usage of data-flow matrices in the simulator. Therefore, the price of this low modeling error is reduced simulation performance in comparison to conventional models. The data-flow matrices contribute to the 2$\times$ reduced simulator performance in comparison to Noxim. However, we strongly advocate this feature. We are convinced that the reduced simulator performance is a price well-payed for the modeling error. The only other viable option to get such good results are bit-accurate simulations. However, those have much worse performance than our data-flow matrices. A trace file of the data transmissions along each link would need to be written. While our data flow matrices have a constant size and a memory complexity of $O(n^2)$, with $n$ colors, generation of bit-accurate trace file has a non-constant size linearly increasing with the simulation time $t$. Since usually the simulation time is much larger than the number of different data streams (i.e.\ colors) in the application graph ($t >> n$), the performance gain and memory reduction are significant.

For the sake of completeness, we also briefly discuss the router architectures properties. As one can see in Fig.~\ref{fig:routerPerf}, the network saturates after 7\% injection rate, which is expected for the chosen, light-weight architecture. Since the focus of this submission is not a novel router architecture but a simulation tool that ships a hardware implementation on top of the actual core simulator, we do not compare against other router implementations. 

To summarize, the \emph{Ratatoskr} framework generates more accurate results than state-of-the-art competitors at the cost of slightly reduced simulation performance. Since accurate power results are key in heterogeneous 3D SoC due to layers in mixed-signal technology with a high base power consumption, \emph{Ratatoskr} tackle one of the most important issues. Furthermore, we ship a hardware implementation, which is automatically generated for each simulation run. Since the whole tool uses a single-point of entry and easy-to-use configuration interfaces it is very user friendly and allows for rapid prototyping. In addition, more detailed configuration interfaces are also provided if non-standard parameters are to be set and optimized. Thus, the proposed design and simulation tool allows to efficiently build NoC for heterogeneous 3D SoCs.

%% file: chapters/05_conclusion.tex
\section{Conclusion}\label{sec:conclusion}

In this work, we introduce \emph{Ratatoskr}, an open-source framework for in-depth PPA analysis in 3D NoCs. We also support heterogeneous 3D integration as it has become one of the key innovations to build more efficient systems. The framework \emph{Ratatoskr} is implemented in C++, SystemC, Python and VHDL. It offers power estimation of routers and links on a cycle-accurate level. The accuracy of our models for the dynamic power of links is within 2.4\% accuracy of bit-level simulations while maintaining cycle-accurate simulation speed. The performance of the NoC can be measured on CA level for long simulations using traffic injected from TL-modeled applications or synthetic patterns, on RTL for shorter simulations using synthetic patterns, and on gate-level for router timing. The hardware implementation of the routers can be synthesized for standard cell technologies and includes a power and area saving feature that removes unused turns in the crossbar based on information about the routing algorithm. This saves up to 32\% total router power and 3\% router area compared to a conventional router without these features. The whole framework evolves around a single configuration file that allows to set the most important design parameters easily, but more detailed and more complex configuration options are also available. The framework generates user reports to assess designs. With this wide range of features, \emph{Ratatoskr} is the first comprehensive framework for PPA-analysis in NoCs also comprising heterogeneous 3D integration. It will participate to tackle important issues for state-of-the-art chips due to the increasing relevance of heterogeneity and the prevalent challenges found in on-chip interconnection networks. You'll find the source code and usage examples of \emph{Ratatoskr} at \underline{\smash{\url{https://github.com/jmjos/ratatoskr}}}.

%% file: main.bbl

\begin{thebibliography}{34}


\ifx \showCODEN    \undefined \def \showCODEN     #1{\unskip}     \fi
\ifx \showDOI      \undefined \def \showDOI       #1{#1}\fi
\ifx \showISBNx    \undefined \def \showISBNx     #1{\unskip}     \fi
\ifx \showISBNxiii \undefined \def \showISBNxiii  #1{\unskip}     \fi
\ifx \showISSN     \undefined \def \showISSN      #1{\unskip}     \fi
\ifx \showLCCN     \undefined \def \showLCCN      #1{\unskip}     \fi
\ifx \shownote     \undefined \def \shownote      #1{#1}          \fi
\ifx \showarticletitle \undefined \def \showarticletitle #1{#1}   \fi
\ifx \showURL      \undefined \def \showURL       {\relax}        \fi
\providecommand\bibfield[2]{#2}
\providecommand\bibinfo[2]{#2}
\providecommand\natexlab[1]{#1}
\providecommand\showeprint[2][]{arXiv:#2}

\bibitem[\protect\citeauthoryear{??}{Int}{[n.d.]}]%
        {Intel.2019}
 \bibinfo{year}{[n.d.]}\natexlab{}.
\newblock \bibinfo{title}{{Intel Previews New Hybrid CPU Architecture with
  Foveros 3D Packaging}}.
\newblock
  \bibinfo{howpublished}{\url{https://newsroom.intel.com/video-archive/video-intel-previews-new-hybrid-cpu-architecture-with-foveros-3d-packaging/}}.
\newblock
\newblock
\shownote{Accessed: 2019-05-17.}


\bibitem[\protect\citeauthoryear{Agarwal, Krishna, Peh, and Jha}{Agarwal
  et~al\mbox{.}}{2009}]%
        {Agarwal.2009}
\bibfield{author}{\bibinfo{person}{Niket Agarwal}, \bibinfo{person}{Tushar
  Krishna}, \bibinfo{person}{Li-Shiuan Peh}, {and} \bibinfo{person}{Niraj~K.
  Jha}.} \bibinfo{year}{2009}\natexlab{}.
\newblock \showarticletitle{{GARNET: A detailed on-chip network model inside a
  full-system simulator}}. In \bibinfo{booktitle}{\emph{{Performance Analysis
  of Systems and Software, 2009. ISPASS 2009. IEEE International Symposium
  on}}}. \bibinfo{pages}{33--42}.
\newblock


\bibitem[\protect\citeauthoryear{{anan-cn}}{{anan-cn}}{2014}]%
        {OpenNoCStanford}
\bibfield{author}{\bibinfo{person}{{anan-cn}}.}
  \bibinfo{year}{2014}\natexlab{}.
\newblock \bibinfo{title}{{Open-Source Network-on-Chip Router RTL}}.
\newblock
  \bibinfo{howpublished}{\url{https://github.com/anan-cn/Open-Source-Network-on-Chip-Router-RTL}}.
\newblock
\newblock
\shownote{Accessed: 2019-03-15.}


\bibitem[\protect\citeauthoryear{Bamberg and Garc{\'i}a-Oritz}{Bamberg and
  Garc{\'i}a-Oritz}{2017}]%
        {Bamberg.2017}
\bibfield{author}{\bibinfo{person}{L. Bamberg} {and} \bibinfo{person}{A.
  Garc{\'i}a-Oritz}.} \bibinfo{year}{2017}\natexlab{}.
\newblock \showarticletitle{{High-Level Energy Estimation for Submicrometric
  TSV Arrays}}.
\newblock \bibinfo{journal}{\emph{{IEEE Transactions on Very Large Scale
  Integration (VLSI) Systems}}} \bibinfo{volume}{25}, \bibinfo{number}{10}
  (\bibinfo{year}{2017}), \bibinfo{pages}{2856--2866}.
\newblock
\showISSN{1063-8210}
\urldef\tempurl%
\url{https://doi.org/10.1109/TVLSI.2017.2713601}
\showDOI{\tempurl}


\bibitem[\protect\citeauthoryear{Bamberg, Joseph, Pionteck, and
  Garcia-Ortiz}{Bamberg et~al\mbox{.}}{2019}]%
        {Bamberg.2018b}
\bibfield{author}{\bibinfo{person}{Lennart Bamberg},
  \bibinfo{person}{Jan~Moritz Joseph}, \bibinfo{person}{Thilo Pionteck}, {and}
  \bibinfo{person}{Alberto Garcia-Ortiz}.} \bibinfo{year}{2019}\natexlab{}.
\newblock \showarticletitle{Crosstalk optimization for through-silicon vias by
  exploiting temporal signal misalignment}.
\newblock \bibinfo{journal}{\emph{Integration}}  \bibinfo{volume}{67}
  (\bibinfo{year}{2019}), \bibinfo{pages}{60 -- 72}.
\newblock
\showISSN{0167-9260}
\urldef\tempurl%
\url{https://doi.org/10.1016/j.vlsi.2019.04.009}
\showDOI{\tempurl}


\bibitem[\protect\citeauthoryear{Bamberg, {Joseph, J. M.}, Schmidt, Pionteck,
  and Garc{\'i}a-Oritz}{Bamberg et~al\mbox{.}}{2018}]%
        {Bamberg.2018}
\bibfield{author}{\bibinfo{person}{L. Bamberg}, \bibinfo{person}{{Joseph, J.
  M.}}, \bibinfo{person}{R. Schmidt}, \bibinfo{person}{T. Pionteck}, {and}
  \bibinfo{person}{A. Garc{\'i}a-Oritz}.} \bibinfo{year}{2018}\natexlab{}.
\newblock \showarticletitle{{Coding-aware Link Energy Estimation for 2D and 3D
  Net\-works-on-Chip with Virtual Channels}}.
\newblock \bibinfo{journal}{\emph{{International Symposium on Power and Timing
  Modeling, Optimization and Simulation}}} (\bibinfo{year}{2018}),
  \bibinfo{pages}{222--228}.
\newblock


\bibitem[\protect\citeauthoryear{Becker}{Becker}{2012}]%
        {Becker.2014}
\bibfield{author}{\bibinfo{person}{Daniel~U. Becker}.}
  \bibinfo{year}{2012}\natexlab{}.
\newblock \bibinfo{booktitle}{\emph{Efficient Microarchitecture for
  Network-on-Chip Routers}}.
\newblock \bibinfo{publisher}{Stanford University}.
\newblock


\bibitem[\protect\citeauthoryear{Becker and Dally}{Becker and Dally}{2009}]%
        {Becker.2009}
\bibfield{author}{\bibinfo{person}{Daniel~U. Becker} {and}
  \bibinfo{person}{William~J. Dally}.} \bibinfo{year}{2009}\natexlab{}.
\newblock \showarticletitle{Allocator Implementations for Network-on-chip
  Routers}. In \bibinfo{booktitle}{\emph{Proceedings of the Conference on High
  Performance Computing Networking, Storage and Analysis}}.
  \bibinfo{publisher}{ACM}, \bibinfo{pages}{1--12}.
\newblock
\urldef\tempurl%
\url{https://doi.org/10.1145/1654059.1654112}
\showDOI{\tempurl}


\bibitem[\protect\citeauthoryear{Catania, Mineo, Monteleone, Palesi, and
  Patti}{Catania et~al\mbox{.}}{2016}]%
        {Catania.2016}
\bibfield{author}{\bibinfo{person}{V. Catania}, \bibinfo{person}{A. Mineo},
  \bibinfo{person}{S. Monteleone}, \bibinfo{person}{M. Palesi}, {and}
  \bibinfo{person}{D. Patti}.} \bibinfo{year}{2016}\natexlab{}.
\newblock \showarticletitle{{Cycle-Accurate Network on Chip Simulation with
  Noxim}}.
\newblock \bibinfo{journal}{\emph{{ACM Transactions on Modeling and Computer
  Simulation}}} \bibinfo{volume}{27}, \bibinfo{number}{1}
  (\bibinfo{year}{2016}), \bibinfo{pages}{1--25}.
\newblock
\urldef\tempurl%
\url{https://doi.org/10.1145/2953878}
\showDOI{\tempurl}


\bibitem[\protect\citeauthoryear{Chen, Emer, and Sze}{Chen
  et~al\mbox{.}}{2018}]%
        {Chen.2018}
\bibfield{author}{\bibinfo{person}{Yu{-}Hsin Chen}, \bibinfo{person}{Joel~S.
  Emer}, {and} \bibinfo{person}{Vivienne Sze}.}
  \bibinfo{year}{2018}\natexlab{}.
\newblock \showarticletitle{Eyeriss v2: {A} Flexible and High-Performance
  Accelerator for Emerging Deep Neural Networks}.
\newblock \bibinfo{journal}{\emph{CoRR}}  \bibinfo{volume}{abs/1807.07928}
  (\bibinfo{year}{2018}).
\newblock
\showeprint[arxiv]{1807.07928}
\urldef\tempurl%
\url{http://arxiv.org/abs/1807.07928}
\showURL{%
\tempurl}


\bibitem[\protect\citeauthoryear{Dong and Xie}{Dong and Xie}{2009}]%
        {Dong.2009}
\bibfield{author}{\bibinfo{person}{X. Dong} {and} \bibinfo{person}{Y. Xie}.}
  \bibinfo{year}{2009}\natexlab{}.
\newblock \showarticletitle{{System-level cost analysis and design exploration
  for three-dimensional integrated circuits (3D ICs)}}.
\newblock \bibinfo{journal}{\emph{{Asia and South Pacific Design Automation
  Conference}}} (\bibinfo{year}{2009}).
\newblock
\urldef\tempurl%
\url{https://doi.org/10.1109/ASPDAC.2009.4796486}
\showDOI{\tempurl}


\bibitem[\protect\citeauthoryear{Dongarra}{Dongarra}{2016}]%
        {Dongarra.2016}
\bibfield{author}{\bibinfo{person}{Jack Dongarra}.}
  \bibinfo{year}{2016}\natexlab{}.
\newblock \bibinfo{booktitle}{\emph{Report on the Sunway TaihuLight System}}.
\newblock \bibinfo{type}{{T}echnical {R}eport}.
  \bibinfo{institution}{University of Tennessee, Oak Ridge National
  Laboratory}.
\newblock
\urldef\tempurl%
\url{http://www.netlib.org/utk/people/JackDongarra/PAPERS/sunway-report-2016.pdf}
\showURL{%
\tempurl}


\bibitem[\protect\citeauthoryear{Drewes, {Joseph, J. M.}, and Pionteck}{Drewes
  et~al\mbox{.}}{2017}]%
        {Drewes.2017}
\bibfield{author}{\bibinfo{person}{T. Drewes}, \bibinfo{person}{{Joseph, J.
  M.}}, {and} \bibinfo{person}{T. Pionteck}.} \bibinfo{year}{2017}\natexlab{}.
\newblock \showarticletitle{{An FPGA-based prototyping framework for
  Networks-on-Chip}}.
\newblock In \bibinfo{booktitle}{\emph{{International Conference on
  ReConFigurable and FPGAs}}}. \bibinfo{publisher}{IEEE},
  \bibinfo{pages}{1--7}.
\newblock
\urldef\tempurl%
\url{https://doi.org/10.1109/RECONFIG.2017.8279775}
\showDOI{\tempurl}


\bibitem[\protect\citeauthoryear{Fatollahi-Fard, Donofrio, Michelogiannakis,
  Shalf, Bachan, and Burke}{Fatollahi-Fard et~al\mbox{.}}{2019}]%
        {OpenSoCFabric}
\bibfield{author}{\bibinfo{person}{F. Fatollahi-Fard}, \bibinfo{person}{D.
  Donofrio}, \bibinfo{person}{G. Michelogiannakis}, \bibinfo{person}{K. Shalf,
  J. adn~Wen}, \bibinfo{person}{J. Bachan}, {and} \bibinfo{person}{D. Burke}.}
  \bibinfo{year}{2019}\natexlab{}.
\newblock \bibinfo{title}{{OpenSoCFabric}}.
\newblock \bibinfo{howpublished}{\url{http://www.opensocfabric.org/home.php}}.
\newblock
\newblock
\shownote{Accessed: 2019-03-15.}


\bibitem[\protect\citeauthoryear{Fazzino, Palesi, and Patti}{Fazzino
  et~al\mbox{.}}{2008}]%
        {Fazzino.2008}
\bibfield{author}{\bibinfo{person}{Fabrizio Fazzino}, \bibinfo{person}{Maurizio
  Palesi}, {and} \bibinfo{person}{David Patti}.}
  \bibinfo{year}{2008}\natexlab{}.
\newblock \showarticletitle{{Noxim: Network-on-chip simulator}}.
\newblock \bibinfo{journal}{\emph{{URL: http://sourceforge.
  net/projects/noxim}}} (\bibinfo{year}{2008}).
\newblock


\bibitem[\protect\citeauthoryear{Garrou, Koyanagi, and Ramm}{Garrou
  et~al\mbox{.}}{2009}]%
        {Garrou.2009}
\bibfield{author}{\bibinfo{person}{P.~E. Garrou}, \bibinfo{person}{M.
  Koyanagi}, {and} \bibinfo{person}{P. Ramm}.} \bibinfo{year}{2009}\natexlab{}.
\newblock \bibinfo{booktitle}{\emph{{3D process technology: Robust circuit and
  physical design for sub-65 nm technology nodes}} (\bibinfo{edition}{first
  edition} ed.)}. \bibinfo{series}{{Handbook of 3D integration}},
  Vol.~\bibinfo{volume}{volume 3}.
\newblock \bibinfo{publisher}{Wiley}, \bibinfo{address}{Hoboken, NJ}.
\newblock
\showISBNx{978-3-527-32034-9}


\bibitem[\protect\citeauthoryear{Jiang, Balfour, Becker, Towles, Dally,
  Michelogiannakis, and Kim}{Jiang et~al\mbox{.}}{2013}]%
        {Jiang.2013}
\bibfield{author}{\bibinfo{person}{Nan Jiang}, \bibinfo{person}{James Balfour},
  \bibinfo{person}{Daniel~U. Becker}, \bibinfo{person}{Brian Towles},
  \bibinfo{person}{William~J. Dally}, \bibinfo{person}{George
  Michelogiannakis}, {and} \bibinfo{person}{John Kim}.}
  \bibinfo{year}{2013}\natexlab{}.
\newblock \showarticletitle{{A detailed and flexible cycle-accurate
  Network-on-Chip simulator}}. In \bibinfo{booktitle}{\emph{{International
  Symposium on Performance Analysis of Systems and Software}}}.
  \bibinfo{publisher}{IEEE}, \bibinfo{pages}{86--96}.
\newblock
\urldef\tempurl%
\url{https://doi.org/10.1109/ISPASS.2013.6557149}
\showDOI{\tempurl}


\bibitem[\protect\citeauthoryear{Jiang, Michelogiannakis, Becker, Towles, and
  Dally}{Jiang et~al\mbox{.}}{[n.d.]}]%
        {Jiang.}
\bibfield{author}{\bibinfo{person}{N. Jiang}, \bibinfo{person}{G.
  Michelogiannakis}, \bibinfo{person}{D. Becker}, \bibinfo{person}{B. Towles},
  {and} \bibinfo{person}{W. Dally}.} \bibinfo{year}{[n.d.]}\natexlab{}.
\newblock \showarticletitle{{Booksim interconnection network simulator}}.
\newblock  (\bibinfo{year}{[n.\,d.]}).
\newblock
\urldef\tempurl%
\url{nocs.stanford.edu}
\showURL{%
\tempurl}


\bibitem[\protect\citeauthoryear{{Joseph}, {Bamberg}, {Ermel}, {Perjikolaei},
  {Drewes}, {García-Ortiz}, and {Pionteck}}{{Joseph} et~al\mbox{.}}{2019}]%
        {Joseph.2018b}
\bibfield{author}{\bibinfo{person}{J.~M. {Joseph}}, \bibinfo{person}{L.
  {Bamberg}}, \bibinfo{person}{D. {Ermel}}, \bibinfo{person}{B.~R.
  {Perjikolaei}}, \bibinfo{person}{A. {Drewes}}, \bibinfo{person}{A.
  {García-Ortiz}}, {and} \bibinfo{person}{T. {Pionteck}}.}
  \bibinfo{year}{2019}\natexlab{}.
\newblock \showarticletitle{NoCs in Heterogeneous 3D SoCs: Co-Design of Routing
  Strategies and Microarchitectures}.
\newblock \bibinfo{journal}{\emph{IEEE Access}}  \bibinfo{volume}{7}
  (\bibinfo{year}{2019}), \bibinfo{pages}{135145--135163}.
\newblock
\showISSN{2169-3536}
\urldef\tempurl%
\url{https://doi.org/10.1109/ACCESS.2019.2942129}
\showDOI{\tempurl}


\bibitem[\protect\citeauthoryear{{Joseph, J. M.}, Bamberg, Krell, Hajjar,
  Garc{\'i}a-Oritz, and Pionteck}{{Joseph, J. M.} et~al\mbox{.}}{2018}]%
        {Joseph.2018}
\bibfield{author}{\bibinfo{person}{{Joseph, J. M.}}, \bibinfo{person}{L.
  Bamberg}, \bibinfo{person}{G. Krell}, \bibinfo{person}{I. Hajjar},
  \bibinfo{person}{A. Garc{\'i}a-Oritz}, {and} \bibinfo{person}{T. Pionteck}.}
  \bibinfo{year}{2018}\natexlab{}.
\newblock \showarticletitle{{Specification of Simulation Models for NoCs in
  Heterogeneous 3D SoCs}}.
\newblock \bibinfo{journal}{\emph{{International Symposium on Reconfigurable
  Communication-centric Systems-on-Chip}}} (\bibinfo{year}{2018}),
  \bibinfo{pages}{1--8}.
\newblock


\bibitem[\protect\citeauthoryear{{Joseph, J. M.}, Blochwitz, Garc{\'i}a-Ortiz,
  and Pionteck}{{Joseph, J. M.} et~al\mbox{.}}{2017}]%
        {Joseph.2017}
\bibfield{author}{\bibinfo{person}{{Joseph, J. M.}}, \bibinfo{person}{C.
  Blochwitz}, \bibinfo{person}{A. Garc{\'i}a-Ortiz}, {and} \bibinfo{person}{T.
  Pionteck}.} \bibinfo{year}{2017}\natexlab{}.
\newblock \showarticletitle{{Area and power savings via asymmetric organization
  of buffers in 3D-NoCs for heterogeneous 3D-SoCs}}.
\newblock \bibinfo{journal}{\emph{{Microprocessors and Microsystems}}}
  \bibinfo{volume}{48} (\bibinfo{year}{2017}), \bibinfo{pages}{36--47}.
\newblock
\showISSN{0141-9331}
\urldef\tempurl%
\url{https://doi.org/10.1016/j.micpro.2016.09.011}
\showDOI{\tempurl}


\bibitem[\protect\citeauthoryear{{Joseph, J. M.} and Pionteck}{{Joseph, J. M.}
  and Pionteck}{2014}]%
        {Joseph.2014}
\bibfield{author}{\bibinfo{person}{{Joseph, J. M.}} {and} \bibinfo{person}{T.
  Pionteck}.} \bibinfo{year}{2014}\natexlab{}.
\newblock \showarticletitle{{A cycle-accurate Network-on-Chip simulator with
  support for abstract task graph modeling}}. In
  \bibinfo{booktitle}{\emph{{International Symposium on System-on-Chip}}}.
  \bibinfo{publisher}{IEEE}, \bibinfo{pages}{1--6}.
\newblock
\showISBNx{978-1-4799-6890-9}
\urldef\tempurl%
\url{https://doi.org/10.1109/ISSOC.2014.6972440}
\showDOI{\tempurl}


\bibitem[\protect\citeauthoryear{{{Joseph, J. M.} co-shared with Bamberg, L.},
  Hajjar, Schmidt, Pionteck, and Garc\'ia-Ortiz}{{{Joseph, J. M.} co-shared
  with Bamberg, L.} et~al\mbox{.}}{2019}]%
        {Joseph.2018c}
\bibfield{author}{\bibinfo{person}{{{Joseph, J. M.} co-shared with Bamberg,
  L.}}, \bibinfo{person}{I. Hajjar}, \bibinfo{person}{R. Schmidt},
  \bibinfo{person}{T. Pionteck}, {and} \bibinfo{person}{A. Garc\'ia-Ortiz}.}
  \bibinfo{year}{2019}\natexlab{}.
\newblock \showarticletitle{{Simulation Environment for Link Energy Estimation
  in Networks-on-Chip with Virtual Channels}}.
\newblock \bibinfo{journal}{\emph{{INTEGRATION, the VLSI journal}}}
  (\bibinfo{year}{2019}).
\newblock


\bibitem[\protect\citeauthoryear{Kahng, Li, Peh, and Samadi}{Kahng
  et~al\mbox{.}}{2009}]%
        {Kahng.2009}
\bibfield{author}{\bibinfo{person}{A.~B. Kahng}, \bibinfo{person}{Bin Li},
  \bibinfo{person}{L.~S. Peh}, {and} \bibinfo{person}{K. Samadi}.}
  \bibinfo{year}{2009}\natexlab{}.
\newblock \showarticletitle{{ORION 2.0: A fast and accurate NoC power and area
  model for early-stage design space exploration}}. In
  \bibinfo{booktitle}{\emph{{Design, Automation Test in Europe Conference
  Exhibition, 2009. DATE '09}}}. \bibinfo{pages}{423--428}.
\newblock
\urldef\tempurl%
\url{https://doi.org/10.1109/DATE.2009.5090700}
\showDOI{\tempurl}


\bibitem[\protect\citeauthoryear{Kahng, Lin, and Nath}{Kahng
  et~al\mbox{.}}{2015}]%
        {Kahng.2015}
\bibfield{author}{\bibinfo{person}{Andrew~B Kahng}, \bibinfo{person}{Bill Lin},
  {and} \bibinfo{person}{Siddhartha Nath}.} \bibinfo{year}{2015}\natexlab{}.
\newblock \showarticletitle{{ORION3.0: a comprehensive NoC router estimation
  tool}}.
\newblock \bibinfo{journal}{\emph{IEEE Embedded Systems Letters}}
  \bibinfo{volume}{7}, \bibinfo{number}{2} (\bibinfo{year}{2015}),
  \bibinfo{pages}{41--45}.
\newblock


\bibitem[\protect\citeauthoryear{Katti, Stucchi, de~Meyer, and Dehaene}{Katti
  et~al\mbox{.}}{2010}]%
        {Katti.2010}
\bibfield{author}{\bibinfo{person}{G. Katti}, \bibinfo{person}{M. Stucchi},
  \bibinfo{person}{K. de Meyer}, {and} \bibinfo{person}{W. Dehaene}.}
  \bibinfo{year}{2010}\natexlab{}.
\newblock \showarticletitle{{Electrical modeling and characterization of
  through silicon via for three-dimensional ICs}}.
\newblock \bibinfo{journal}{\emph{{IEEE Transactions on Electron Devices}}}
  \bibinfo{volume}{57}, \bibinfo{number}{1} (\bibinfo{year}{2010}),
  \bibinfo{pages}{256--262}.
\newblock
\urldef\tempurl%
\url{https://doi.org/10.1109/TED.2009.2034508}
\showDOI{\tempurl}


\bibitem[\protect\citeauthoryear{Kim, Lee, Kim, Kim, and Yoo}{Kim
  et~al\mbox{.}}{2009}]%
        {Kim.2009}
\bibfield{author}{\bibinfo{person}{K. Kim}, \bibinfo{person}{S. Lee},
  \bibinfo{person}{J.~Y. Kim}, \bibinfo{person}{M. Kim}, {and}
  \bibinfo{person}{H.~J. Yoo}.} \bibinfo{year}{2009}\natexlab{}.
\newblock \showarticletitle{{A 125 GOPS 583 mW Network-on-Chip Based Parallel
  Processor With Bio-Inspired Visual Attention Engine}}.
\newblock \bibinfo{journal}{\emph{{IEEE Journal of Solid-State Circuits}}}
  \bibinfo{volume}{44}, \bibinfo{number}{1} (\bibinfo{year}{2009}),
  \bibinfo{pages}{136--147}.
\newblock
\showISSN{0018-9200}
\urldef\tempurl%
\url{https://doi.org/10.1109/JSSC.2008.2007157}
\showDOI{\tempurl}


\bibitem[\protect\citeauthoryear{Krishna and Kwon}{Krishna and Kwon}{2017}]%
        {OpenSmart}
\bibfield{author}{\bibinfo{person}{K. Krishna} {and} \bibinfo{person}{H.
  Kwon}.} \bibinfo{year}{2017}\natexlab{}.
\newblock \bibinfo{title}{{OpenSMART}}.
\newblock
  \bibinfo{howpublished}{\url{http://synergy.ece.gatech.edu/tools/opensmart/}}.
\newblock
\newblock
\shownote{Accessed: 2019-03-15.}


\bibitem[\protect\citeauthoryear{{Kwon} and {Krishna}}{{Kwon} and
  {Krishna}}{2017}]%
        {Kwon.2017}
\bibfield{author}{\bibinfo{person}{H. {Kwon}} {and} \bibinfo{person}{T.
  {Krishna}}.} \bibinfo{year}{2017}\natexlab{}.
\newblock \showarticletitle{OpenSMART: Single-cycle multi-hop NoC generator in
  BSV and Chisel}. In \bibinfo{booktitle}{\emph{2017 IEEE International
  Symposium on Performance Analysis of Systems and Software (ISPASS)}}.
  \bibinfo{pages}{195--204}.
\newblock
\urldef\tempurl%
\url{https://doi.org/10.1109/ISPASS.2017.7975291}
\showDOI{\tempurl}


\bibitem[\protect\citeauthoryear{Lepak, Talbot, White, Beck, and
  Naffziger}{Lepak et~al\mbox{.}}{2017}]%
        {Lepak.2017}
\bibfield{author}{\bibinfo{person}{K. Lepak}, \bibinfo{person}{G. Talbot},
  \bibinfo{person}{S. White}, \bibinfo{person}{N. Beck}, {and}
  \bibinfo{person}{S. Naffziger}.} \bibinfo{year}{2017}\natexlab{}.
\newblock \showarticletitle{The next generation AMD enterprise server product
  architecture}. In \bibinfo{booktitle}{\emph{Hotchips 29}}.
\newblock


\bibitem[\protect\citeauthoryear{Markov}{Markov}{2014}]%
        {Markov.2014}
\bibfield{author}{\bibinfo{person}{I.~L. Markov}.}
  \bibinfo{year}{2014}\natexlab{}.
\newblock \showarticletitle{{Limits on fundamental limits to computation}}.
\newblock \bibinfo{journal}{\emph{{Nature}}} (\bibinfo{year}{2014}).
\newblock


\bibitem[\protect\citeauthoryear{Penolazzi and Jantsch}{Penolazzi and
  Jantsch}{2006}]%
        {Penolazzi.2006}
\bibfield{author}{\bibinfo{person}{S. Penolazzi} {and} \bibinfo{person}{A.
  Jantsch}.} \bibinfo{year}{2006}\natexlab{}.
\newblock \showarticletitle{{A High Level Power Model for the Nostrum NoC}}. In
  \bibinfo{booktitle}{\emph{{9th EUROMICRO Conference on Digital System Design:
  Architectures, Methods and Tools}}}, \bibfield{editor}{\bibinfo{person}{Venki
  Muthukumar}} (Ed.). \bibinfo{publisher}{{IEEE Computer Society}},
  \bibinfo{address}{Los Alamitos, Calif.}, \bibinfo{pages}{673--676}.
\newblock
\showISBNx{0-7695-2609-8}
\urldef\tempurl%
\url{https://doi.org/10.1109/DSD.2006.9}
\showDOI{\tempurl}


\bibitem[\protect\citeauthoryear{Yu, Li, Zhang, Pan, and He}{Yu
  et~al\mbox{.}}{2013}]%
        {Yu.2013}
\bibfield{author}{\bibinfo{person}{X. Yu}, \bibinfo{person}{L. Li},
  \bibinfo{person}{Y. Zhang}, \bibinfo{person}{H. Pan}, {and}
  \bibinfo{person}{S. He}.} \bibinfo{year}{2013}\natexlab{}.
\newblock \showarticletitle{{Performance and power consumption analysis of
  memory efficient 3D network-on-chip architecture}}.
\newblock \bibinfo{journal}{\emph{{International Conference on Control and
  Automation}}} (\bibinfo{year}{2013}).
\newblock
\urldef\tempurl%
\url{https://doi.org/10.1109/ICCA.2013.6565107}
\showDOI{\tempurl}


\bibitem[\protect\citeauthoryear{Zar{\'a}ndy}{Zar{\'a}ndy}{2011}]%
        {Zarandy.2011}
\bibfield{author}{\bibinfo{person}{{\'A}. Zar{\'a}ndy}.}
  \bibinfo{year}{2011}\natexlab{}.
\newblock \bibinfo{booktitle}{\emph{{Focal-plane sensor-processor chips}}}.
\newblock \bibinfo{publisher}{Springer}.
\newblock
\showISBNx{9781441964755}


\end{thebibliography}
